\documentclass[12pt]{article}
\usepackage[utf8]{inputenc}
\usepackage[DIV=13]{typearea}\usepackage{ragged2e}
\usepackage{calligra,amsmath,amsfonts,bbm,mathrsfs,amssymb}
\usepackage{mathrsfs}
\usepackage{slashed,cancel}
\usepackage{cite}
\usepackage{hyperref}
\usepackage{cleveref}
\hypersetup{colorlinks=true,urlcolor=magenta,anchorcolor=blue,citecolor=blue,filecolor=blue,linkcolor=magenta,menucolor=blue,linktocpage=true,pdfproducer=medialab,pdfauthor={Joel Jones, Luca Merlo, Javier Silva Malpartida},pdftitle={XXX}}
\usepackage[english]{babel}
\usepackage{indentfirst}
\usepackage{graphicx}
\usepackage{float}
\usepackage{enumerate}
\usepackage{caption}
\usepackage{subcaption}
\usepackage{multirow}
\usepackage{xcolor}
\usepackage{booktabs}
\usepackage{physics}
\usepackage{url}
\usepackage{soul}
\usepackage{verbatim}
\usepackage{multicol}
\usepackage{bbold}
\usepackage{fontawesome}
\usepackage{tablefootnote}
\usepackage[normalem]{ulem}
\usepackage{bm}
\usepackage{sectsty}
\allsectionsfont{\boldmath}
\newcommand{\email}[1]{\href{mailto:#1}{\tt #1}}

\numberwithin{equation}{section}

\newcommand{\blue}[1]{\color{blue} #1 \color{black}}
\newcommand{\magenta}[1]{\color{magenta} #1 \color{black}}
\newcommand{\be}{\begin{equation}}
\newcommand{\ee}{\end{equation}}
\newcommand{\ba} {\begin{equation}\begin{aligned}}
\newcommand{\ea} {\end{aligned}\end{equation}}
\newcommand{\bea}{\begin{eqnarray}}
\newcommand{\eea}{\end{eqnarray}}

\newcommand{\sL}{\mathscr{L}}

\newcommand{\hc}{\text{h.c.}}
\renewcommand{\sL}{\mathscr{L}}

\newcommand{\mcL}{\mathcal{L}}

\newcommand\subsetsim{\mathrel{
  \ooalign{\raise0.2ex\hbox{$\subset$}\cr\hidewidth\raise-0.8ex\hbox{\scalebox{0.9}{$\sim$}}\hidewidth\cr}}}

\newcommand{\TeV}{\ \text{TeV}}
\newcommand{\GeV}{\ \text{GeV}}
\newcommand{\MeV}{\ \text{MeV}}

\usepackage{color}
\usepackage{colortbl}
\definecolor{rossoc}{cmyk}{0,1,1,0.2}
\definecolor{dgreen}{rgb}{0.0, 0.5, 0.0}

\begin{document} 
\renewcommand*{\thefootnote}{\fnsymbol{footnote}}

\begin{titlepage}

\vspace*{-1cm}
\flushleft{\magenta{IFT-UAM/CSIC-26-136}} 
\\[1cm]

\begin{center}
\blue{\bf \Large ALP-Mediated Dark Matter at Low Reheating Temperature and the Belle~II Excess}\\
\centering
\vskip .3cm
\end{center}
\vskip 0.5  cm
\begin{center}
{\large\bf Joel Jones-Pérez}$^{a}$~\footnote{\email{jones.j@pucp.edu.pe}},
{\large\bf Luca Merlo}$^{b}$~\footnote{\email{luca.merlo@uam.es}}, and
{\large\bf Javier Silva-Malpartida}$^{a}$~\footnote{\email{javier.silvam@pucp.edu.pe}},
\vskip .7cm
{\footnotesize
$^a$~Secci\'on F\'isica, Departamento de Ciencias, Pontificia Universidad Cat\'olica del Per\'u, \\ Apartado 1761, Lima, Per\'u
{\par\centering \vskip 0.25 cm\par}
$^b$~Departamento de F\'isica Te\'orica and Instituto de F\'isica Te\'orica UAM/CSIC,\\
Universidad Aut\'onoma de Madrid, Cantoblanco, 28049, Madrid, Spain
}

\end{center}
\vskip 1cm

\begin{abstract}
\justify

Belle~II has reported a $2.7\sigma$ excess in $B^+\to K^+\nu\bar\nu$ over its Standard Model
prediction, recently interpreted as arising from an axion-like particle mediating a thermal dark
matter candidate. We reassess this interpretation using a new mass-differential measurement of the
same data set and adding two visible-decay searches that earlier analyses omitted. The resulting
bounds leave the predictions from a standard cosmological history only marginally compatible with the excess, viable
in a narrow window of parameter space. Thus, we consider a situation where 
dark matter freezes out during a low-temperature reheating era, featuring entropy injection diluting the relic abundance. This requires smaller couplings compared to those in standard cosmology in order to match the observed dark matter density, opening a region of parameter space where the compatibility of the model with the Belle~II excess is significantly improved. 
\end{abstract}
\end{titlepage}
\setcounter{footnote}{0}

\pdfbookmark[1]{Table of Contents}{tableofcontents}
\tableofcontents
\renewcommand*{\thefootnote}{\arabic{footnote}}

\bigskip
\section{Introduction}
\label{sec:intro}

Dark matter (DM) is one of our surest signs indicating the need for an extension of the Standard
Model (SM). Since the theoretical result, dated to 1977, that a stable particle in thermal equilibrium
with the SM plasma freezes out with an abundance set by its annihilation cross
section~\cite{Lee:1977ua,Vysotsky:1977pe,Hut:1977zn}, the weakly interacting massive particle
(WIMP) has been considered a very attractive DM candidate (see Ref.~\cite{Arcadi:2024ukq}
for a recent review). However, direct and indirect experimental searches for WIMPs with masses above
$\mathcal{O}(10)\GeV$ have led to null results~\cite{LZ:2024zvo,XENON:2023cxc,Fermi-LAT:2015att},\footnote{Notice there has been recently a $2.6\sigma$ hint for direct detection at the LZ experiment~\cite{LZ:2026axp}.} motivating the consideration of DM
candidates with lower masses, capable of avoiding the aforementioned bounds. Of course, testing such candidates requires the consideration of different search strategies (see for example Refs.~\cite{Bringmann:2018cvk,Wang:2021jic,Gago:2026ocx,Baryakhtar:2020rwy,DallaValleGarcia:2025cwf,Bernal:2025qkj,Arias:2025tvd,Bae:2023ago,Kopp:2016yji}). One particular strategy is to probe their influence in rare decays at high precision experiments.

In this context, the Belle~II experiment has reported evidence for $B^+\to K^+\nu\bar\nu$
decays~\cite{Belle-II:2023esi}, measuring a branching ratio some $2.7\sigma$ above the SM
prediction~\cite{Parrott:2022zte}. Because the experimental signature is a kaon and nothing else,
the excess is described equally well by a two-body decay $B^+\to K^+X$, with $X$ a neutral
particle that is either stable, decays to invisible states, or is long-lived enough to escape the
detector~\cite{Altmannshofer:2023hkn}. This naturally calls for an interpretation of the anomaly
in terms of a mediator coupling to a $(\bar b\,\Gamma s)$ operator and, should that mediator also
couple to the dark sector, for a direct link between the excess and DM. Thus, in this work, we explore the possibility of having an axion-like particle (ALP) as such a mediator.

ALPs are among the best motivated mediators between DM and the SM. They arise as
pseudo Nambu-Goldstone bosons of a spontaneously broken global symmetry, as in the original
solution of the strong CP problem~\cite{Peccei:1977hh,Weinberg:1977ma,Wilczek:1977pj}, and appear
in a wide range of ultraviolet scenarios. They have been considered as DM candidates in their own
right~\cite{Gelmini:1984pe},
their impact on cosmological observables has been investigated in
detail especially in the last years ~\cite{Ferreira:2018vjj,DEramo:2018vss,Escudero:2019gvw,Arias-Aragon:2020qtn,Arias-Aragon:2020qip,Arias-Aragon:2020shv,Ferreira:2020bpb},
and they remain an evergreen topic in flavour model
building~\cite{Davidson:1981zd,Wilczek:1982rv,Ema:2016ops,Calibbi:2016hwq,Arias-Aragon:2017eww,Arias-Aragon:2022ats,DiLuzio:2023ndz,Greljo:2024evt} and neutrino mass generation mechanisms~\cite{Chikashige:1980qk,Chikashige:1980ui,Gelmini:1980re,Biggio:2023gtm,deGiorgi:2023tvn,Greljo:2025suh,Debnath:2026ebg,Biggio:2026gcs}.
Recent reviews covering ALP models and their phenomenology are
Refs.~\cite{Albertus:2026fbe,Arza:2026rsl}. This has motivated an extensive experimental programme spanning astrophysics, cosmology, colliders~\cite{Jaeckel:2012yz,Mimasu:2014nea,Knapen:2016moh,Brivio:2017ije,Bauer:2017nlg,Mariotti:2017vtv,Bauer:2017ris,Bauer:2018uxu,deGiorgi:2022oks,Bonilla:2022pxu,Calibbi:2022izs,Marcos:2024yfm,Arias-Aragon:2024gpm,Biekotter:2025fll} and low-energy facilities~\cite{Izaguirre:2016dfi,Marciano:2016yhf,Merlo:2019anv,Aloni:2019ruo,Bauer:2019gfk,Cornella:2019uxs,Calibbi:2020jvd,Bauer:2021mvw,Guerrera:2021yss,Gallo:2021ame,Bonilla:2022qgm,Bonilla:2022vtn,Guerrera:2022ykl,DiLuzio:2024jip,deGiorgi:2024str,Arias-Aragon:2024gdz,Bisht:2024hbs,Calibbi:2024rcm,MartinCamalich:2025srw,Alda:2025uwo,Alda:2025nsz,Ardu:2026vsr,Zamoro:2026ily}. 

The ALP mass and its couplings to DM and SM fields are independent parameters, allowing it to be a convenient portal to the dark sector. In this way, a single particle controls both how the dark sector is populated in the early Universe, and how it may be seen in the laboratory.

Two recent analyses have made the connection between the Belle~II excess and ALP-mediated DM explicit, considering
opposite ends of the coupling range. Ref.~\cite{Ding:2025eqq} follows the original idea in Ref.~\cite{Dolan:2017osp} to explore the resonant
annihilation, $\chi\chi\to a^*\to\gamma\gamma$, to reproduce the
observed abundance through freeze-out. On the other hand, Ref.~\cite{Calibbi:2025rpx} populates the dark sector
via freeze-in. Each rests on a different projection of the same Belle~II events, with the former
using the measured inclusive rate as a target band~\cite{Belle-II:2023esi}, and the latter based on a dedicated
reinterpretation~\cite{Fridell:2023ssf}.

Since these results were published, a mass-differential likelihood for the same data set has since become available~\cite{Abumusabh:2025zsr}. Here, a finer binning resolves a narrow two-body signal that coarser projections dilute, so that the resulting constraint on ${\rm Br}(B^+\to K^+a)$ has to be evaluated mass by mass rather than as a single inclusive rate. This motivates performing a reanalysis of ALP-mediated scenarios from the flavour point of view. Thus, as a primary objective of this work, the analysis of Ref.~\cite{Abumusabh:2025zsr} will be adopted as a sole Belle~II input, and the collider viability of the ALP-mediated DM scenario will be recomputed.

A missing-energy measurement on its own is, however, not enough to test this scenario. Given that ALP couplings to SM fields are necessary to explain Belle~II, ALP decays into visible particles must be included into its phenomenology. Since missing-energy searches lose their sensitivity around the point where visible ones acquire theirs, we add two constraints that neither earlier analyses imposed: the LHCb search for $B^+\to K^+a(\to\mu^+\mu^-)$~\cite{LHCb:2016awg}, and the Belle search for the prompt diphoton final state $B\to K^{(*)}a(\to\gamma\gamma)$~\cite{Belle:2025bkgammagamma}. These searches probe the ALP directly, rather than through its absence, and will be essential to constrain the regions of the parameter space where invisible decay is either forbidden or suppressed.

Regarding DM, the analyses above, Refs.~\cite{Ding:2025eqq,Dolan:2017osp,Calibbi:2025rpx}, assume a standard cosmological history. The evolution of the Universe
before Big Bang Nucleosynthesis (BBN) is, however, untested~\cite{Allahverdi:2020bys}. In fact, the
reheating at the end of inflation may have completed as late as a few MeV, with the minimum reheating temperature determined by BBN and
neutrino decoupling~\cite{Kawasaki:2000en,Hannestad:2004px,deSalas:2015glj,Bernal:2018kcw,Bernal:2022wck,Drees:2006vh}. Here, the entropy
injected into the SM plasma as the inflaton decays dilutes the abundance previously generated by freeze-out, so that
matching the observed relic density requires a smaller annihilation rate and therefore smaller
couplings. This can also be the case if, alternatively, an early matter-dominated era may have intervened~\cite{Drees:2017iod}. Either
possibility alters the relation between the couplings of a DM candidate and its relic abundance,
opening regions of parameter space that standard cosmology closes~\cite{Silva-Malpartida:2023yks,Silva-Malpartida:2024emu}. It is then of interest to test if a modified cosmology can give us further insights on how our model can solve the Belle~II anomaly, simultaneously reproducing the observed relic abundance. In particular, we intend to evaluate if it is possible to achieve this without recurring to the somewhat tuned freeze-out funnel region of the parameter space.

To summarize, the purpose of this work is to reassess the ALP-mediated DM interpretation of the
$B^+\to K^++{\rm inv}$ excess against the latest experimental input, and to ask whether the DM can
still be a thermal relic once the assumption of a standard cosmological history is dropped. The
flavour and collider analysis is performed with \texttt{ALPaca}~\cite{Alda:2025nsz}.
Sec.~\ref{sec:alpportal} fixes the Belle~II input, defines the model and works out the terrestrial
constraints, invisible and visible, that follow from $B$ decays.
Sec.~\ref{sec:cosmo} turns to the early Universe, and confronts the relic abundance obtained under
a low reheating temperature with those constraints. Our conclusions follow.

\section[\boldmath An ALP interpretation of $B^+\to K^+ + {\rm inv}$\unboldmath]{An ALP interpretation of $B^+\to K^+ + {\rm inv}$}
\label{sec:alpportal}

Belle~II reported the first evidence for $B^+\to K^+\nu\bar\nu$~\cite{Belle-II:2023esi},
\be
{\rm Br}(B^+\to K^+\nu\bar\nu) =
\left(2.3\pm0.5\,({\rm stat})\,^{+0.5}_{-0.4}\,({\rm syst})\right)\times10^{-5}\,,
\label{eq:belle2}
\ee
some $2.7\sigma$ above the SM prediction of $(5.58\pm0.37)\times10^{-6}$~\cite{Parrott:2022zte}. Ref.~\cite{Altmannshofer:2023hkn} reinterpreted the measurement in terms of a two-body decay
$B^+\to K^+a$, fitting 12 inclusive-tag-analysis (ITA) cells obtained from three coarse $q^2$ bins
in each of four signal-discriminator intervals and obtaining a non-zero best-fit rate for an ALP mass around $2\GeV$, rather than
an upper limit.

Furthermore, as mentioned in the Introduction, Ref.~\cite{Abumusabh:2025zsr} has recently released a mass-differential likelihood for the same Belle~II dataset, built from the published ITA and hadronic-tag-analysis (HTA) spectra, to fit simultaneously the continuous $B^+\to K^+\nu\bar\nu$ component and a narrow $B^+\to K^+ a$ signal. The reconstructed ITA spectrum has sufficiently fine $q^2_{\rm rec}$ binning to resolve a narrow resonance, and is supplemented by the nearly orthogonal HTA sample. It therefore provides a mass-dependent constraint on ${\rm Br}(B^+\to K^+a)$ rather than a single inclusive rate. In particular, a nearly massless mediator cannot account for the excess, while for a GeV-scale ALP the limit varies appreciably across mass bins, having to be evaluated point by point. It is of interest, therefore, to reinterpret the feasibility of our ALP-mediated DM model considering this analysis as our nominal input.\footnote{Notice there exist other analyses of the excess, see Refs~\cite{Gartner:2026clx,Abumusabh:2026ykp}. Nevertheless, we consider the fine physical-variable projection of~\cite{Abumusabh:2025zsr} to be better suited to a narrow two-body signal.}

Our study intends to be broad, not focusing on $B\to K+{\rm inv}$ alone, as a complete phenomenology must also consider other ALP production processes, as well as ALP decays into different channels, both of which can be constrained by other searches. 
Among the ``invisible" alternatives, the BaBar collaboration~\cite{BaBar:2013npw} provides two important constraints than must be taken into consideration. First comes $B\to K^*+{\rm inv}$, which is governed by the same $b\to s\,a$ coefficient, with different form factors and kinematics. Here, we use the same BaBar dataset and find that, within our region of interest, $B\to K^*a$ reaches at most a $\sim1\sigma$ discrepancy, against $2.6\sigma$ for $B\to Ka$. It can therefore be safely neglected as an additional bound.
The second constraint from BaBar comes from an earlier $B^+\to K^+\nu\bar\nu$ search. We have checked explicitly that the sensitivity of BaBar is far weaker than that of Ref.~\cite{Abumusabh:2025zsr} throughout the mass range considered here, so can be omitted without affecting the results.

We do find that the ``visible" LHCb dimuon search for $B^+\to K^+a(\to\mu^+\mu^-)$~\cite{LHCb:2016awg} as well the dedicated search for the prompt diphoton final state $B\to K^{(*)}a(\to\gamma\gamma)$~\cite{Belle:2025bkgammagamma} can put important bounds on our parameter space. In particular, they cover the region below the dark-matter threshold that $B\to K+{\rm inv}$ cannot constrain at all, as well as constrain the regions where visible ALP decays dominate over the invisible ones.

Before proceeding with our analysis, let us comment that Ref.~\cite{Ding:2025eqq} considered the same ALP-mediated DM model, treating the raw measured rate in Eq.~\eqref{eq:belle2} directly as a range where to fit the $B^+\to K^+a$ branching ratio. Requiring the same ALP to mediate resonant dark-matter annihilation, $\chi\chi\to a^*\to\gamma\gamma$ near $m_a\simeq2m_\chi$, and to reproduce the observed relic abundance, Ref.~\cite{Ding:2025eqq} identified a viable region in their parameter space. However, our final results based on the updated analysis of Ref.~\cite{Abumusabh:2025zsr}, combined with the visible-channel bounds, disfavour this interpretation such that, even though some points of the model can still lead to the correct relic density, every explicit benchmark point of Ref.~\cite{Ding:2025eqq} is excluded.

We proceed as follows. We define the model in Sec.~\ref{sec:model} and collect the rates entering these invisible and visible searches in
Secs.~\ref{sec:formulae} and~\ref{sec:formulae-visible}, respectively. In Sec.~\ref{sec:combined-terrestrial} we present the main results of our analysis. Finally, in Sec.~\ref{sec:rhoEW-outlook} we study the consequences of relaxing a key assumption taken in Ref.~\cite{Ding:2025eqq}.

\subsection{The ALP and the dark sector}
\label{sec:model}

Below the scale at which the global symmetry is broken, the interactions of an ALP with the SM
fields are well captured by an effective field theory (EFT) expansion in inverse powers of the ALP
scale $f_a$, a description that the community has extensively worked to improve and
complete~\cite{Choi:1986zw,Salvio:2013iaa,Brivio:2017ije,Alonso-Alvarez:2018irt,Gavela:2019wzg,Chala:2020wvs,DiLuzio:2020oah,Bonilla:2021ufe,Arias-Aragon:2022byr,Arias-Aragon:2022iwl,DiLuzio:2023cuk,DiLuzio:2023lmd}.
We take the ALP $a$ to be a gauge-singlet pseudoscalar whose only contact with the SM is
through the electroweak gauge bosons. Above the electroweak scale, following the conventions of Ref.~\cite{Ding:2025eqq}, we have:
\ba
\mcL_{\rm EW} = \frac{1}{2}\,\partial^\mu a\,\partial_\mu a - \frac{1}{2}\,m_a^2\,a^2
- \frac{c_B}{4 f_a}\,a\,B^{\mu\nu}\widetilde{B}_{\mu\nu}
- \frac{c_W}{4 f_a}\,a\,W^{i,\mu\nu}\widetilde{W}^i_{\mu\nu}\,,
\label{eq:LEW}
\ea
where $B^{\mu\nu}$ and $W^{i,\mu\nu}$ are the $U(1)_Y$ and $SU(2)_L$ field strengths, the duals
are defined as $\widetilde{X}_{\mu\nu}\equiv\frac{1}{2}\epsilon_{\mu\nu\rho\sigma}X^{\rho\sigma}$,
and the ALP mass $m_a$ and decay constant $f_a$ are treated as independent
parameters.\footnote{The dimensionless coefficients $c_B$ and $c_W$ carry no explicit loop
factor in this normalisation. The numerical results of this section are obtained
with \texttt{ALPaca}~\cite{Alda:2025nsz}, whose
\texttt{derivative\_above} basis factors out $g_i^2/16\pi^2$, so that
$c_B=-(\alpha_{em}/\pi\cos^2\theta_W)\,c_B^{\tt ALPaca}$ and
$c_W=-(\alpha_{em}/\pi\sin^2\theta_W)\,c_W^{\tt ALPaca}$. The benchmark of
Eq.~\eqref{eq:benchmark} below therefore reads
$c_B^{\tt ALPaca}=\cot^2\theta_W\,c_W^{\tt ALPaca}\simeq3.3\,c_W^{\tt ALPaca}$ in that basis.} The latter is a genuine feature of a broader class of axions that solve the Strong CP problem as recently discussed in Refs.~\cite{DiLuzio:2016sbl,DiLuzio:2017pfr,Gaillard:2018xgk,Hook:2019qoh,DiLuzio:2020wdo,DiLuzio:2020oah,DiLuzio:2021pxd,DiLuzio:2021gos,Gavela:2023tzu,Cox:2023dou,deGiorgi:2024elx,FernandezNavarro:2026cyu}. The EFT expansion itself only requires $m_a$ below the cutoff $4\pi f_a$, for a strongly interacting underlying theory, or the cutoff $f_a$, for a weakly interacting ultraviolet realisation. 

Since our working hypothesis is that the same particle accounts for the Belle~II excess in
$B^+\to K^++{\rm inv}$, the two-body kinematics of $B^+\to K^+a$ restricts the ALP to be lighter than $m_B-m_K\simeq4.79\GeV$, and produced with appreciable phase space. We accordingly take $m_a$ between a few hundred MeV and the kinematic endpoint throughout.

After electroweak symmetry breaking, the two anomalous operators in Eq.~\eqref{eq:LEW} generate couplings
of the ALP to all four combinations of neutral and charged electroweak gauge bosons,
\ba
\sL_{\rm EW} \supset
-\frac{g_{a\gamma\gamma}}{4}\,a\,F_{\mu\nu}\widetilde{F}^{\mu\nu}
-\frac{g_{a\gamma Z}}{4}\,a\,F_{\mu\nu}\widetilde{Z}^{\mu\nu}
-\frac{g_{aZZ}}{4}\,a\,Z_{\mu\nu}\widetilde{Z}^{\mu\nu}
-\frac{g_{aWW}}{4}\,a\,W_{\mu\nu}\widetilde{W}^{\mu\nu}\,,
\label{eq:LEWSB}
\ea
of which only the photon and $W$ couplings matter for our purposes. In terms of the coefficients of Eq.~\eqref{eq:LEW},
\be
g_{a\gamma\gamma} = \frac{c_B\cos^2\theta_W + c_W\sin^2\theta_W}{f_a}\,,
\qquad
g_{aWW} = \frac{c_W}{f_a}\,,
\label{eq:gagg-gaWW}
\ee
with $\theta_W$ the weak mixing angle.

The dark sector consists of a Majorana fermion singlet $\chi$, coupled to the ALP through the
dimension-five derivative operator
\ba
\sL_{\rm DM} = \bar{\chi}\,i\slashed{\partial}\,\chi - m_\chi\,\bar{\chi}\chi
- g_{a\chi\chi}\,\partial_\mu a\,\bar{\chi}\gamma^\mu\gamma_5\chi\,,
\label{eq:LDM}
\ea
where $g_{a\chi\chi}$ has dimensions of inverse mass. Thus, the model is specified by five independent parameters: $m_a$,
$m_\chi$, $g_{aWW}$, $g_{a\gamma\gamma}$, and $g_{a\chi\chi}$. The coupling to $W$ bosons $g_{aWW}$ generates the flavour-changing $b\to s\,a$ vertex and therefore
governs how the ALP is produced in $B$ decays. The coupling to photons $g_{a\gamma\gamma}$ governs how the ALP decays back into visible states. As we have seen, the two descend from the same pair of ultraviolet coefficients and are independent in general.
The third coupling, $g_{a\chi\chi}$, ties the ALP derivatively to $\chi$.
This coupling controls the invisible width $a\to\chi\chi$ and, together with $g_{a\gamma\gamma}$, the annihilation
$\chi\chi\to a^*\to\gamma\gamma$ that determines the relic abundance. Whether the ALP decays into DM or
back into photons is decided by the interplay of $g_{a\gamma\gamma}$ and $g_{a\chi\chi}$, being crucial for the flavour phenomenology which follows.

In most of our analysis, we shall follow the benchmark of Ref.~\cite{Ding:2025eqq} and take $c_B=c_W$ as a simplifying assumption, such that the two gauge coefficients are equal:
\be
g_{a\gamma\gamma}=g_{aWW}\equiv g_a\,.
\label{eq:benchmark}
\ee
In this way we have one coupling governing both the production of the ALP in $B$ decays and its decay back into
visible states. We will explore the consequences of breaking this relation in Sec.~\ref{sec:rhoEW-outlook}.

Eq.~\eqref{eq:LEW} does not contain tree-level coupling of the  ALP to fermions or to gluons, so its decays
into SM states are controlled exclusively by $g_{aWW}$ and $g_{a\gamma\gamma}$~\cite{Alda:2025uwo}. The diphoton partial width is
\be
\Gamma(a\to\gamma\gamma) = \frac{g_{a\gamma\gamma}^2\,m_a^3}{64\pi}\,,
\label{eq:Gagg}
\ee
and, when the channel is kinematically open, the partial width into dark matter is
\be
\Gamma(a\to\chi\chi) = \frac{g_{a\chi\chi}^2\,m_a\,m_\chi^2}{\pi}
\sqrt{1-\frac{4m_\chi^2}{m_a^2}}\,.
\label{eq:Gachi}
\ee

Charged-lepton and hadronic final states arise only from gauge-boson loops and from the running
between the ultraviolet and the electroweak scale, since the ALP has no tree-level coupling to
leptons, quarks or gluons. Due to this radiative origin, these couplings lie many orders of magnitude below
the diphoton channel. For illustration, Table~\ref{tab:BRvis} collects the branching ratios into
SM states across the mass range of interest, as calculated by \texttt{ALPaca}~\cite{Alda:2025nsz} at the benchmark of Eq.~(\ref{eq:benchmark}),
 assuming that invisible decays are closed. 

\begin{table}[t]
\centering
\begin{tabular}{lccccc}
\toprule
$m_a$ [GeV] & $\gamma\gamma$ & $e^+e^-$ & $\mu^+\mu^-$ & $\tau^+\tau^-$ & hadrons \\
\midrule
$0.30$ & $0.999986$ & $2.1\times10^{-8}$  & $1.4\times10^{-5}$ & -- & $4.2\times10^{-15}$ \\
$0.50$ & $0.999989$ & $1.0\times10^{-8}$  & $1.1\times10^{-5}$ & -- & $2.8\times10^{-10}$ \\
$1.00$ & $0.999994$ & $3.8\times10^{-9}$  & $5.7\times10^{-6}$ & -- & $2.2\times10^{-7}$  \\
$2.00$ & $0.999997$ & $1.3\times10^{-9}$  & $2.4\times10^{-6}$ & -- & $2.1\times10^{-7}$  \\
$2.65$ & $0.999993$ & $9.3\times10^{-10}$ & $1.7\times10^{-6}$ & -- & $5.3\times10^{-6}$  \\
$3.55$ & $0.999989$ & $5.6\times10^{-10}$ & $1.1\times10^{-6}$ & -- & $9.6\times10^{-6}$  \\
$4.30$ & $0.999976$ & $4.2\times10^{-10}$ & $8.9\times10^{-7}$ & $1.4\times10^{-5}$ & $9.2\times10^{-6}$ \\
\bottomrule
\end{tabular}
\caption{\it Branching ratios of the ALP into SM final states at the benchmark of
Eq.~\eqref{eq:benchmark}, normalised to the total SM width. A dash denotes a closed channel;
$\tau^+\tau^-$ opens at $m_a=2m_\tau\simeq3.554\GeV$, just above the $3.55\GeV$ entry. The
hadronic column is inclusive, and switches description within the range shown, from exclusive
light-meson modes below $m_a\simeq2.3\GeV$ to $a\to gg$ and, above $m_a\simeq2.55\GeV$,
$a\to c\bar{c}$; the charm threshold makes it rise by more than an order of magnitude within a
few tens of MeV, so the tabulated values should not be interpolated across it.}
\label{tab:BRvis}
\end{table}

Among the hadronic channels, parity forbids a pseudoscalar from decaying into two pseudoscalars,
so $a\to\pi\pi$ is closed and the lightest hadronic states are reached through three-body
channels. The widths computed by \texttt{ALPaca}~\cite{Alda:2025nsz} include
$a\to3\pi$, $a\to\pi\pi\gamma$, $a\to\eta\pi\pi$ and $a\to\eta'\pi\pi$, as well as $a\to\omega\omega$
where it is open. These are evaluated in chiral perturbation theory for $m_a\lesssim1\GeV$ and corrected
above by resonance contributions extracted from data. The switch to a perturbative
quark-hadron-duality description at higher $m_a$, visible in the last rows of the Table, is
detailed in the caption.

From the Table, we clearly see that the ALP is a diphoton resonance to better than one part in $10^4$ throughout, with every
other SM channel at the level of a few times $10^{-5}$ or below.

Let us now compare the diphoton and invisible widths. These scale differently, the diphoton channel (Eq.~\eqref{eq:Gagg}) grows as $m_a^3$ and as $g_{a\gamma\gamma}^2$, while the invisible channel (Eq.~\eqref{eq:Gachi}) 
depends on $g_{a\chi\chi}^2$, being linear in $m_a$ and $m_{\chi}^2$ in the limit $m_{\chi}^2\ll m_a^2$. This difference will drive the phenomenology of the next Sections. The invisible branching ratio then reads
\be
{\rm Br}(a\to{\rm inv}) = \frac{\Gamma(a\to\chi\chi)}
{\Gamma(a\to\chi\chi)+\Gamma_{\rm SM}}\,,
\label{eq:Brinv}
\ee
where $\Gamma_{\rm SM}$ is the total width into SM final states. At the benchmark of
Eq.~\eqref{eq:benchmark}, Table~\ref{tab:BRvis} shows that
$\Gamma_{\rm SM}\simeq\Gamma(a\to\gamma\gamma)$ to the accuracy relevant here. 
The invisible
branching ratio is therefore close to one only in the region of the parameter space where $\Gamma(a\to\chi\chi)\gg\Gamma_{\rm SM}$. In the opposite regime it falls to
\be
{\rm Br}(a\to{\rm inv})
\;\xrightarrow[\ \Gamma_{\gamma\gamma}\gg\Gamma_{\chi\chi}\ ]{}\;
64\,\left(\frac{g_{a\chi\chi}}{g_{a\gamma\gamma}}\right)^2\,\left(\frac{m_\chi}{m_a}\right)^2\,\beta_\chi\,,
\label{eq:Brinv-limit}
\ee 
with $\beta_\chi=\sqrt{1-4m_\chi^2/m_a^2}$. The invisible fraction vanishes identically below the threshold $m_a=2m_\chi$, where the ALP has no
dark final state to decay into and can only appear as missing energy if it is long-lived enough to
leave the detector.

\subsection{Rates for $B\to K^{(*)}a$}
\label{sec:formulae}

Below the electroweak scale, the $c_W$ operator of Eq.~\eqref{eq:LEW} induces a flavour-changing
coupling of the ALP to left-handed quark currents through $W$ exchange (see Ref.~\cite{Alda:2025uwo} for a recent review). For the $b\to s$
transition the effective coupling is
\be
\mcL_{\rm eff} \supset -\,g_{aWW}\,C_{sb}\;\partial^\mu a\;
\left(\bar{s}_L\gamma_\mu b_L\right) + \hc\,,
\label{eq:Leff-bsa}
\ee
with the Wilson coefficient generated at one loop and dominated by the top quark,
\be
C_{sb} = -\frac{3\,m_W^2}{16\pi^2 v^2}\,V_{ts}^*V_{tb}\;
f\!\left(\frac{m_t^2}{m_W^2}\right)\,,
\qquad
f(x) = \frac{x\left[1+x(\log x-1)\right]}{(1-x)^2}\,.
\label{eq:Csb}
\ee
One can see that Eq.~\eqref{eq:Leff-bsa} contains a vector and an axial-vector current in equal measure, which ties the pseudoscalar and vector kaon final states together such that they are not independent: $B\to Ka$ is controlled by the vector component and $B\to K^*a$ by the axial-vector one. The two-body widths then read
\begin{align}
\Gamma(B^\pm\to K^\pm a) &= \frac{g_{aWW}^2\,|C_{sb}|^2}{64\pi\,m_B^3}\,
f_0^2(m_a^2)\,\left(m_B^2-m_K^2\right)^2\,
\lambda^{1/2}\!\left(m_B^2,m_K^2,m_a^2\right)\,,
\label{eq:GBKa}\\[2mm]
\Gamma(B\to K^* a) &= \frac{g_{aWW}^2\,|C_{sb}|^2\,m_B^3}{64\pi}\,
A_0^2(m_a^2)\,
\lambda^{3/2}\!\left(1,\frac{m_{K^*}^2}{m_B^2},\frac{m_a^2}{m_B^2}\right)\,,
\label{eq:GBKstara}
\end{align}
where
$\lambda(\alpha,\beta,\gamma)=\alpha^2+\beta^2+\gamma^2-2(\alpha\beta+\alpha\gamma+\beta\gamma)$
is the K\"all\'en function, and $f_0$ and $A_0$ are the scalar and pseudoscalar $B\to K^{(*)}$
form factors evaluated at $q^2=m_a^2$. We take $f_0^{B\to K}$ from the lattice average
of Ref.~\cite{FlavourLatticeAveragingGroupFLAG:2021npn} and $A_0^{B\to K^*}$ from
Ref.~\cite{Horgan:2015vla}, as implemented in \texttt{flavio}~\cite{Straub:2018kue}. Both rise
steeply towards the endpoint of the decay, such that over the mass range of interest we have
\be
\begin{aligned}
f_0^{B\to K}(m_a^2) &= 0.33,\ 0.37,\ 0.40,\ 0.50,\ 0.66\,,\\[2mm]
A_0^{B\to K^*}(m_a^2) &= 0.37,\ 0.47,\ 0.58,\ 0.90,\ 1.66\,,
\end{aligned}
\label{eq:ffvalues}
\ee
at $m_a=0,\ 2,\ 2.65,\ 3.55$ and $4.3\GeV$ respectively.
The different powers of $\lambda$ in Eqs.~(\ref{eq:GBKa}) and~(\ref{eq:GBKstara}) are worth noting: the $P$-wave $K^*$ mode is suppressed by two
extra powers of the ALP momentum, so it closes faster as $m_a$ approaches the kinematic
endpoint, so it is the weaker of the two constraints over most of the range of interest even
before the experimental sensitivities are folded in.

What the missing-energy experiments actually measure is the product of production and invisible
decay. Following the standard long-lived-particle treatment~\cite{Dolan:2017osp}, we have
\be
{\rm Br}\!\left(B\to K^{(*)}+{\rm inv}\right) =
{\rm Br}\!\left(B\to K^{(*)}a\right)\,
\left[{\rm Br}(a\to\chi\chi)
+ e^{-L_{\rm det}/\beta\gamma\,c\tau_a}\,{\rm Br}(a\to{\rm vis})\right]\,,
\label{eq:BRinv-full}
\ee
where the second term accounts for an ALP that decays visibly but only after leaving the detector,
with $L_{\rm det}$ the detector size and $\tau_a$ and $\beta\gamma$ being the ALP lifetime and boost. Within the parameter ranges scanned in this work, we find the ALP to be short-lived, so the new contribution to ${\rm Br}\!\left(B\to K^{(*)}+{\rm inv}\right)$
reduces to the product of Eq.~\eqref{eq:GBKa} with the invisible branching ratio of Eq.~\eqref{eq:Brinv}.

With this, for each point of the parameter space we compare our predicted branching ratio in Eq.~(\ref{eq:BRinv-full}) with the fit result extracted from Fig.~2 in Ref.~\cite{Abumusabh:2025zsr}. This comparison, to be shown in Section~\ref{sec:combined-terrestrial}, will be presented in terms of the significance $Z_{\rm inv}$, as calculated by \texttt{ALPaca}, with larger $Z_{\rm inv}$ indicating a stronger tension with the fit, as usual.

It is worth noting that, throughout this work, we use $Z_\text{obs}$ for every search to denote a significance in the usual
Gaussian sense: the number of standard deviations by which a point in parameter space is
disfavoured. Where a search reports a central value and its (possibly asymmetric) uncertainty, as
for the missing-energy likelihood of Ref.~\cite{Abumusabh:2025zsr} ($Z_{\rm inv}$, mentioned above) and the
LHCb dimuon search (${Z_{K\mu\mu}}$, below), $Z_\text{obs}$ is the ordinary one-sided Gaussian pull between
our predicted rate and that central value. For the diphoton case, to be discussed in the next section, only a single quoted confidence-level
limit is available, so no such pull can be formed directly, and we instead build a monotonic proxy calibrated to reproduce the stated confidence level of that limit.

\subsection{Visible-channel constraints}
\label{sec:formulae-visible}

Two visible-channel observables complement the missing-energy search of
Sec.~\ref{sec:formulae}, both reconstructed from the same $B\to K^{(*)}a$ production: a dimuon
resonance search and a diphoton resonance search.

The first is an LHCb search for $B^+\to K^+a$ followed by
$a\to\mu^+\mu^-$~\cite{LHCb:2016awg}, which sets limits at $95\%$ confidence level (CL) on:
\be
{\rm Br}\!\left(B^+\to K^+\mu^+\mu^-\right)_a =
{\rm Br}\!\left(B^+\to K^+a\right)\,{\rm Br}\!\left(a\to\mu^+\mu^-\right)\,
\epsilon(m_a,\tau_a)\,,
\label{eq:BRkmumu}
\ee
where $\epsilon$ is the lifetime-dependent efficiency of the search. The bound exists only
above $m_a\simeq0.26\GeV$, where $a\to\mu^+\mu^-$ opens, and carries genuine gaps where LHCb
vetoes dimuon-resonance regions. It probes the sub-threshold band $m_a<2m_\chi$, where
${\rm Br}(a\to{\rm inv})$ vanishes identically and $B\to K+{\rm inv}$ has no prompt-invisible
sensitivity. Above threshold, the opening of the invisible channel does not switch this bound off, although
${\rm Br}(a\to\mu^+\mu^-)$ is now suppressed by a factor $(1-{\rm Br}(a\to{\rm inv}))$. 
This constraint can remain very strong well above threshold, particularly at small
$g_{a\chi\chi}$, where the invisible width stays subdominant over a wide range of $m_a$. 
In the following, the tension between the branching ratio in Eq.~(\ref{eq:BRkmumu}) and the upper limits shown in Fig.~4 of~\cite{LHCb:2016awg} will be quantified by the significance $Z_{K\mu\mu}$.

Another relevant visible observable is the diphoton resonance, based on a joint search for $B\to K^{(*)}a$, $a\to\gamma\gamma$ by Belle/Belle~II, targeting a prompt diphoton final state~\cite{Belle:2025bkgammagamma}.
The search covers $m_a$ from $0.16$ to $4.50\GeV$ for the $K$ modes and sets $90\%$ CL
limits on $g_{aWW}$ at the level of $3\times10^{-6}$--$3\times10^{-5}\,{\rm GeV}^{-1}$, under the
assumption of a purely $W$-induced coupling, $c_B=0$. We find this search to
dominate over the dimuon one throughout essentially the whole parameter space scanned here. The
two swap roles only inside the $\pi^0/\eta/\eta'$ veto windows of the search (around
$m_a\simeq0.50$ and $0.97\GeV$, see Sec.~\ref{sec:combined-terrestrial}), where the dimuon search is the only visible-channel constraint left.
Following the same convention as for the $B^+\to K^++{\rm inv}$ and $B^+\to K^+\mu^+\mu^-$ searches, we quote $B\to K^{(*)}\gamma\gamma$ at $2\sigma$, denoting the limit as $g^{\rm Belle}_{aW,2\sigma}(m_a)$. Since the rate scales as $g_{aWW}^2$, this means we need to rescale the results in Fig.~4 of Ref.~\cite{Belle:2025bkgammagamma} by $\sqrt{2/1.28}\simeq1.25$.

Nevertheless, recasting this limit onto our benchmark is not a simple rescaling of couplings, for two reasons.
First, the signal model from Belle assumes a purely $W$-induced coupling ($c_B=0$), fixing $g_{a\gamma\gamma}^{\rm Belle}=\sin^2\theta_W\,g_{aWW}$ via Eq.~\eqref{eq:gagg-gaWW},
a factor $\sin^2\theta_W$ below
our benchmark $g_{a\gamma\gamma}=g_{aWW}$ of Eq.~\eqref{eq:benchmark}. Second, the limit quoted by Belle is not a
bare coupling bound: it includes an efficiency loss from the finite ALP lifetime,
through a per-mass-hypothesis fit
$\varepsilon(c\tau)/\varepsilon(0)=r_f\,e^{a_1c\tau}+(1-r_f)e^{a_2c\tau}$ with parameters
$a_1,a_2,r_f$ not tabulated in the publication. 
A faithful recast at our benchmark would therefore
require the ratio $\varepsilon_{\rm ours}(c\tau)/\varepsilon_{\rm Belle}(c\tau)$ between our assumed
efficiency at a specific lifetime and that taken by Belle, and unfortunately
this ratio cannot be reconstructed from Ref.~\cite{Belle:2025bkgammagamma} alone.

Since both models share the same
production vertex and detector, the exact exclusion condition is
\be
g_{aWW}\sqrt{{\rm Br}(a\to\gamma\gamma)}\,
\sqrt{\varepsilon_{\rm ours}(c\tau)/\varepsilon_{\rm Belle}(c\tau)}
\ \ge\ g^{\rm Belle}_{aW,2\sigma}(m_a)\,.
\label{eq:belle-diphoton-exclusion}
\ee
Since we are unable to compute the efficiency ratio, we use instead a conservative proxy by setting it equal to unity, and exclude those points satisfying the following relation:
\be
{\cal R}_{\gamma\gamma}\equiv
\frac{g_{aWW}\sqrt{{\rm Br}(a\to\gamma\gamma)}}{g^{\rm Belle}_{aW,2\sigma}(m_a)}\ \ge\ 1\,,
\label{eq:belle-diphoton-proxy}
\ee
with ${\rm Br}(a\to\gamma\gamma)$ evaluated at our benchmark. All in all, we define the corresponding exclusion significance, in units of standard deviations, as $Z_{\gamma\gamma}\equiv2\,{\cal R}_{\gamma\gamma}^2$, so that Eq.~\eqref{eq:belle-diphoton-proxy} is equivalently an exclusion
boundary at $Z_{\gamma\gamma}=2$, i.e.\ at $2\sigma$. 

It is worth noting that this proxy is conservative
rather than merely approximate. Because the diphoton partial width into diphotons scales as $g_{a\gamma\gamma}^2$, the
$\sin^2\theta_W$ suppression of the coupling, identified above, means that at any fixed $g_{aWW}$
this width exceeds the one assumed in the signal model of Belle by
$1/\sin^4\theta_W\simeq19$. Since the total width also includes the dark width $\Gamma(a\to\chi\chi)$, the former will always be larger than that assumed by Belle~II. This 
implies a more promptly decaying, more efficiently reconstructed ALP, 
such that a point excluded by our proxy in fact satisfies
$\varepsilon_{\rm ours}(c\tau)/\varepsilon_{\rm Belle}(c\tau)\ge1$. 
Eq.~\eqref{eq:belle-diphoton-proxy} can therefore never
exclude a point that an exact recast of Eq.~\eqref{eq:belle-diphoton-exclusion} would allow, meaning it can be 
considered as a rigorously conservative bound

From Eq.~\eqref{eq:Brinv-limit}, the ratio $\Gamma(a\to\gamma\gamma)/\Gamma(a\to\chi\chi)$ falls as
$g_{a\chi\chi}^{-2}$. A larger dark coupling therefore diverts an increasing share of ALP decays
into $\chi\chi$, suppressing ${\rm Br}(a\to\gamma\gamma)$. The diphoton search is comparatively
weak wherever the dark coupling is sizeable, precisely where the missing-energy constraint of
Sec.~\ref{sec:formulae} is most constraining. It only becomes competitive with $B\to K\mu\mu$
where $g_{a\chi\chi}$ is small enough that the visible width is not swamped by the dark one.
Sec.~\ref{sec:combined-terrestrial} shows this
pattern directly.

Neither Ref.~\cite{Ding:2025eqq} nor Ref.~\cite{Calibbi:2025rpx} impose neither
the $B^+\to K^+\mu^+\mu^-$ veto of Eq.~\eqref{eq:BRkmumu} nor the $B\to K^{(*)}\gamma\gamma$ constraint of Eq.~\eqref{eq:belle-diphoton-proxy}.
In particular, Ref.~\cite{Ding:2025eqq} restricts
its collider input to missing-energy and other diphoton/kaon searches, while
Ref.~\cite{Calibbi:2025rpx} accounts for visible ALP decays only through a generic
detector-acceptance factor, without recasting an actual dimuon or diphoton exclusion. Both
visible-channel bounds introduced in this Section, $B^+\to K^+a(\to\mu^+\mu^-)$ and $B\to K^{(*)}a(\to\gamma\gamma)$, are genuinely new relative to that prior work.

\subsection{Combined invisibles and visible searches}
\label{sec:combined-terrestrial}

\begin{figure}[t]
\centering
\includegraphics[width=\textwidth]{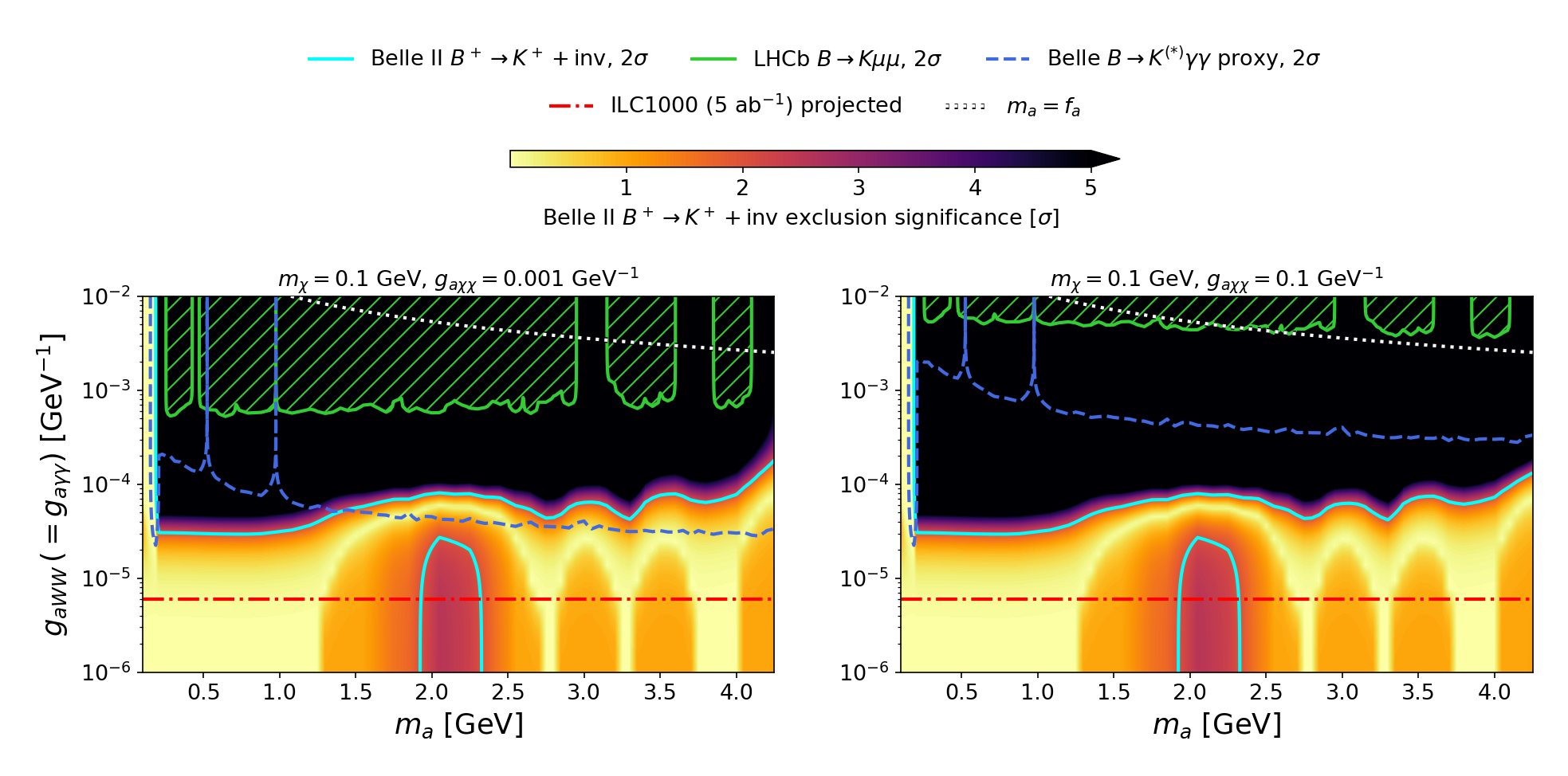}
\caption{\it Missing-energy exclusion significance from Ref.~\cite{Abumusabh:2025zsr} (colour map and cyan
$2\sigma$ contour), with the LHCb $B\to K\mu\mu$ bound (green, hatched: excluded at $>2\sigma$),
the Belle $B\to K^{(*)}a(\gamma\gamma)$ prompt-rate proxy of Eq.~\eqref{eq:belle-diphoton-proxy}
(blue dashed, $2\sigma$) and the $m_a=f_a$ EFT cutoff (white dotted) overlaid, for $m_\chi=0.1\GeV$ and
$g_{a\chi\chi}=10^{-3},10^{-1}\GeV^{-1}$ (left to right). The red dash-dotted line corresponds to the projected ILC
sensitivity~\cite{Bhattacharya:2025hme}.}
\label{fig:combined-terrestrial-mchi01}
\end{figure}

\begin{figure}[t]
\centering
\includegraphics[width=\textwidth]{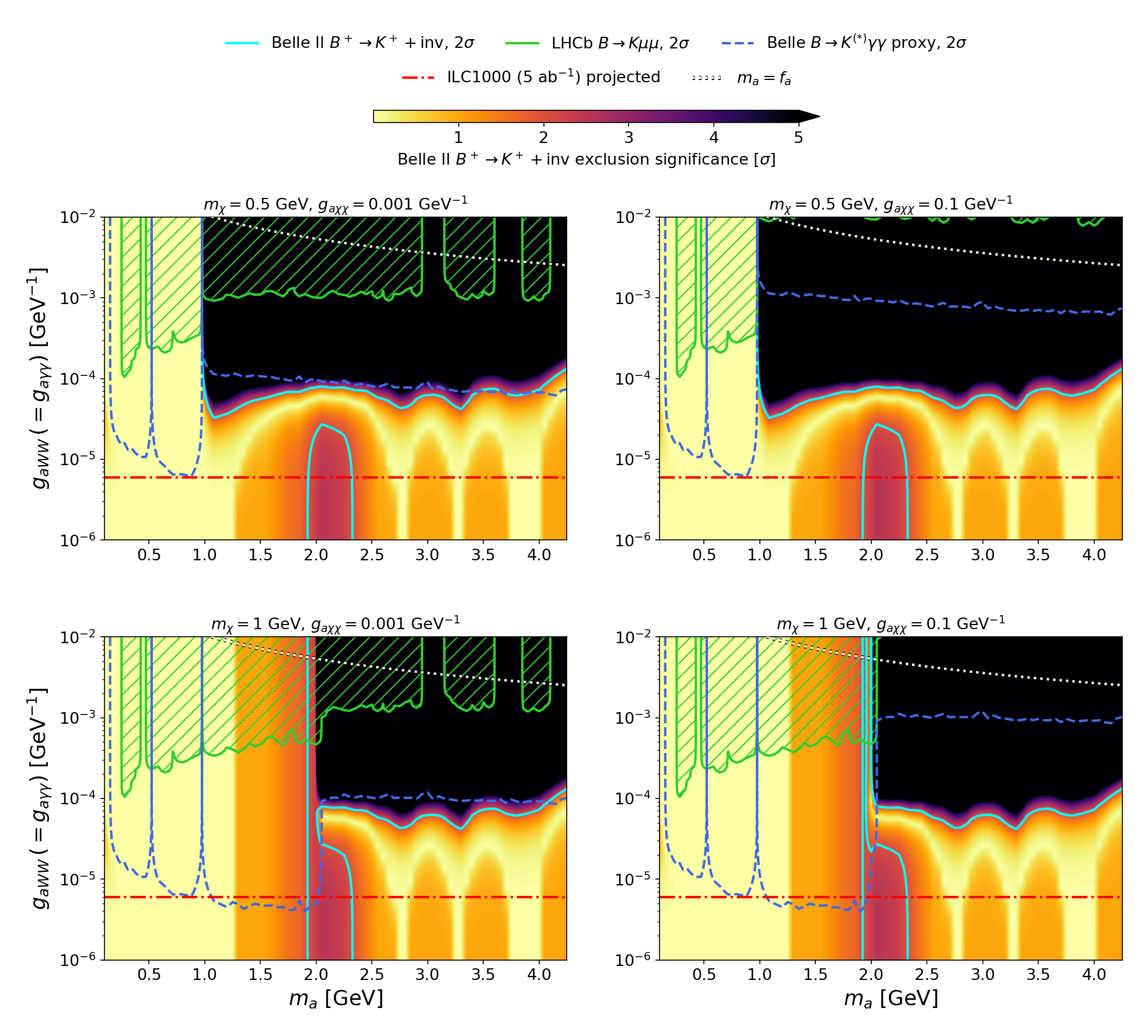}
\caption{\it Same as Fig.~\ref{fig:combined-terrestrial-mchi01}, for $m_\chi=0.5\GeV$ (top row)
and $m_\chi=1\GeV$ (bottom row).}
\label{fig:combined-terrestrial-mchi05-1}
\end{figure}

The three searches introduced above, that is, the missing-energy likelihood of Ref.~\cite{Abumusabh:2025zsr},
the LHCb $B^+\to K^+a(\to\mu^+\mu^-)$ veto of Eq.~\eqref{eq:BRkmumu}, and the
Belle $B\to K^{(*)}a(\to\gamma\gamma)$ proxy of Eq.~\eqref{eq:belle-diphoton-proxy}, probe
disjoint final states of the same underlying process and can therefore be combined without
double-counting. Figs.~\ref{fig:combined-terrestrial-mchi01}--\ref{fig:combined-terrestrial-mchi05-1}
show these constraints overlaid for representative benchmark points considered throughout this work, $m_\chi=0.1,\,0.5,\,1\GeV$ and
$g_{a\chi\chi}=10^{-3},\,10^{-1}\GeV^{-1}$. Here, the significance $Z_{\rm inv}$ is represented
as a colour map, with its $2\sigma$ boundary as a cyan contour. Moreover, the LHCb
$Z_{K\mu\mu}=2$ boundary is shown as a green, hatched contour. Finally, the Belle $Z_{\gamma\gamma}=2$ boundary
of Eq.~\eqref{eq:belle-diphoton-proxy} appears as a blue dashed contour. In addition to this information, the white dotted line marks the
weakly coupled EFT-validity boundary $m_a=f_a$.

Two features of these contours are artefacts of the individual searches rather than genuine
structure in the model. The narrow gaps in the LHCb contour are the dimuon-resonance vetoes
already noted in Sec.~\ref{sec:formulae-visible}; the similar narrow features in the diphoton
contour from Belle at $m_a\simeq0.50$ and $0.97\GeV$ play the same role.
In addition, the $Z_{\rm inv}$ heatmap itself
is shown only up to $m_a=4.3\GeV$, short of the true kinematic endpoint
$m_B-m_K\simeq4.79\GeV$. The reason for this is that, as $m_a\to m_B-m_K$, the two-body phase space in $B\to Ka$ closes and
the bound extracted from the reinterpretation in Ref.~\cite{Abumusabh:2025zsr} becomes numerically
unstable. 
This is consistent with the sensitivity degradation already visible
in their own analysis from $m_a\sim4\GeV$ onwards.

In Fig.~\ref{fig:combined-terrestrial-mchi01}, valid for $m_\chi=0.1$~GeV, we find that the $B\to K+{\rm inv}$ search excludes couplings $g_{aWW}$ above $\mathcal{O}(10^{-5})$~GeV$^{-1}$ across our entire range of ALP mass, with the exclusion weakening at larger $m_a$. A noticeable feature lies at $m_a\approx2$~GeV, where couplings below $\sim3\times10^{-5}$~GeV$^{-1}$ are also excluded, pointing in the same direction as the fit performed in Ref.~\cite{Altmannshofer:2023hkn}. Furthermore, we find that only the diphoton $B\to K^{(*)}a(\to\gamma\gamma)$ search constrains the parameter space favoured by the Belle~II excess, and only for $g_{a\chi\chi}=10^{-3}$~GeV$^{-1}$. In this specific case, we find that the diphoton data rules out large part of the $m_a\approx2$~GeV region.

The Figure also shows a red dash-dotted line, which marks the projected sensitivity $g_{a\gamma\gamma}\simeq6\times10^{-6}\GeV^{-1}$ of a future
$e^+e^-\to\gamma\,a(\to\chi\chi)$ search at a $\sqrt{s}=1\,\TeV$ linear collider with
$5\,{\rm ab}^{-1}$ and polarised beams,
essentially flat over the mass range shown~\cite{Bhattacharya:2025hme}. At
the benchmark of Eq.~\eqref{eq:benchmark}, it maps directly onto the $g_{aWW}$ axis and would
extend the region excluded by diphoton searches by around 1-2 orders of magnitude if realised, being able to further probe this anomaly if strengthened by future data.

Fig.~\ref{fig:combined-terrestrial-mchi05-1} shows the same information for larger values of $m_\chi$. These plots show the consequences of having $m_a<2m_{\chi}$, which closes the $a\to\chi\chi$ invisible decay channel. In this situation, the visible channels dominate the decay, and are crucial in the exclusion of the parameter space. In particular, the diphoton search can rule out couplings as small as $\mathcal{O}(10^{-6})$~GeV$^{-1}$, in some cases being more sensitive than the ILC sensitivity displayed by the red dash-dotted line. Nevertheless, for $m_a>2m_\chi$, the visible searches do not exclude much of the parameter space favoured by the Belle~II excess, leaving the $m_a\approx2$~GeV region untouched.

Let us now consider the results of Ref.~\cite{Ding:2025eqq} in this context. Assuming the ALP-mediated dark matter to be thermal in the early Universe, and reproducing the relic abundance via freeze-out through a resonance, the authors identified a viable region $m_a\in(0.6,\,4.8)\GeV$ with $m_\chi=0.477\,m_a$, $g_{a\gamma\gamma}=g_{aWW}\in(7.13$--$9.60)\times10^{-5}\GeV^{-1}$ and $g_{a\chi\chi}\in(7.12\times10^{-5}$--$7.73\times10^{-3})\GeV^{-1}$. These were expressed in terms of three explicit benchmark points at $m_a=0.75,\,2.65$ and $3.55\GeV$.

Via an extensive analysis including values of $g_{a\chi\chi}$ in addition to those shown in Figs.~\ref{fig:combined-terrestrial-mchi01} and~\ref{fig:combined-terrestrial-mchi05-1}, we find that combining our three searches excludes the three explicit benchmark points of
Ref.~\cite{Ding:2025eqq}. Indeed, both edges of the coupling range at each of $m_a=0.75,\,2.65$ and $3.55\GeV$ are usually ruled out by the invisible search, with the diphoton search excluding the rest. The standard cosmology window of Ref.~\cite{Ding:2025eqq}
survives only where ${\rm Br}(a\to{\rm inv})$ falls short of the near-unity value
their fit assumed.

To make this combination precise, we define at each point of parameter space
\be
q_{\rm excl} = Z_{\rm inv}^2 + Z_{K\mu\mu}^2 + Z_{\gamma\gamma}^2\,, \qquad
Z_{\rm comb}\equiv\sqrt{q_{\rm excl}}\,,
\label{eq:combined-exclusion}
\ee
the significance of each search added in quadrature, and take $Z_{\rm comb}=2$ as the working
exclusion boundary. On this statistic, all six benchmarks points in Ref.~\cite{Ding:2025eqq}
(both coupling edges at each of $m_a=0.75,\,2.65$ and $3.55\GeV$) are excluded at
$Z_{\rm comb}>2.8$.\footnote{Note that Ref.~\cite{Ding:2025eqq} does not itself compute a comparable combined statistic: their
Belle~II ``signal region'' is a band around the measured central value for the branching ratio within its quoted
uncertainty, and every other constraint is simply overlaid with no joint significance of their own. So when we say $Z_{\rm comb}$ excludes their benchmark
points, this is a statement in our combined-significance framework with respect to points on the parameter space, and not a discrepancy with a
comparable number of theirs.}

 \subsection{Beyond the benchmark $c_B=c_W$}
\label{sec:rhoEW-outlook}

Everything above is evaluated at the benchmark of Eq.~\eqref{eq:benchmark}. More generally, the
relation between the two electroweak coefficients is fixed by
\be
\rho_{\rm EW} \equiv \frac{g_{a\gamma\gamma}}{g_{aWW}}\,,
\label{eq:rhoEWdef-main}
\ee
with $\rho_{\rm EW}=1$ the benchmark studied so far. Furthermore, $\rho_{\rm EW}=\sin^2\theta_W\simeq0.23$
corresponds to a vanishing hypercharge coefficient, and $\rho_{\rm EW}\gg1$ to a photophilic
ALP. Another possible benchmark, $\rho_{\rm EW}\to0$, corresponds to a photophobic ALP, which we do not
consider further as it is not relevant for our cosmological analysis.
Thus, for completeness we recomputed the exclusion statistic of Eq.~\eqref{eq:combined-exclusion} at
$\rho_{\rm EW}=0.23$ and $5$, on the same region of the parameter space considered above, to check which features
of Sec.~\ref{sec:combined-terrestrial} survive.

Two effects can be traced analytically. The missing-energy boundary is controlled almost
everywhere by $\Gamma(a\to\chi\chi)$, which does not depend on $\rho_{\rm EW}$, so its position is
essentially insensitive to this choice. The $B\to K\mu\mu$ boundary,
by contrast, cannot be tracked analytically. It
depends on the competition between the loop-induced leptonic width and
$\Gamma(a\to\gamma\gamma)$, both of which respond to $\rho_{\rm EW}$, and has to be evaluated
numerically at each value.

Concretely, in the six panels of
Figs.~\ref{fig:combined-terrestrial-mchi01}--\ref{fig:combined-terrestrial-mchi05-1}, the three
contours respond differently to $\rho_{\rm EW}$. The missing-energy boundary is essentially
unchanged, moving by less than a few percent almost everywhere. At the couplings where it crosses
$Z_{\rm inv}=2$ the ALP already decays almost entirely into $\chi\chi$. Since $\rho_{\rm EW}$ only
enters through the diphoton width, $\Gamma(a\to\gamma\gamma)\propto g_{a\gamma\gamma}^2=\rho_{\rm
EW}^2\,g_{aWW}^2$, changing it has almost no effect on a decay that is already dominated by
$\Gamma(a\to\chi\chi)$. We find that the boundary only weakens, by up to a factor of two, in the
corner of small $g_{a\chi\chi}$ and large $m_a$, where $\Gamma(a\to\gamma\gamma)$ starts to compete
with $\Gamma(a\to\chi\chi)$. The Belle diphoton boundary follows instead the clean scaling
$g_{aWW}\propto\rho_{\rm EW}^{-1/2}$, inherited from $\Gamma(a\to\gamma\gamma)$ as just discussed.
The bound strengthens to smaller couplings by a factor $\simeq2.1$ at $\rho_{\rm EW}=0.23$, and weakens to larger couplings by a factor
$\simeq0.45$ at $\rho_{\rm EW}=5$, uniformly across the mass range and both evaluated values of
$g_{a\chi\chi}$. The LHCb dimuon boundary moves in the same direction by a comparable but
non-universal factor of order $1.5$--$2$; at $g_{a\chi\chi}=10^{-1}\GeV^{-1}$ it can disappear
from the plotted window altogether at $\rho_{\rm EW}\le1$ and only reappear at $\rho_{\rm EW}=5$,
but always deep inside the region the missing-energy contour already excludes, consistent with
the competition of Eq.~\eqref{eq:Brinv-limit}. None of these shifts changes which
of the three searches dominates in a given corner of the grid.

Numerically, the combined invisible and visible exclusion boundary moves by no
more than a factor of a few across $\rho_{\rm EW}\in[0.23,5]$ at fixed $m_\chi,g_{a\chi\chi}$,
markedly less than the shift in $\rho_{\rm EW}$ itself. This means the overall picture is not
qualitatively sensitive to this choice, and the split of Sec.~\ref{sec:combined-terrestrial} between weak- and
strong-dark-coupling regimes persists unchanged. The surviving regions are shown in detail in Sec.~\ref{sec:cosmo-compat}.

\section{The Belle~II excess in front of late reheating}
\label{sec:cosmo}

Secs.~\ref{sec:model}--\ref{sec:rhoEW-outlook} fixed the ALP's couplings to photons, $W$ bosons and dark matter purely from
flavour and collider data, without asking whether the dark sector they define can actually be
populated in the early Universe. In this Section, we address whether $\chi$ can
be a thermal relic, and under what cosmological history, once its interactions are fixed as
in Sec.~\ref{sec:alpportal}. We proceed in three steps. In Sec.~\ref{sec:reheating} we set up the
cosmological history we consider beyond the standard case, a reheating era that may complete at a
temperature as low as 4~MeV, following the limit by BBN. In Sec.~\ref{sec:dmgenesis} we review how this history modifies the freeze-out of $\chi$ relative to the standard calculation. Finally, in Sec.~\ref{sec:cosmo-compat} we overlay the resulting solutions for the relic abundance on the flavour-based exclusion maps of Sec.~\ref{sec:alpportal}, to identify in what way the non-standard cosmology improves the compatibility of the observed dark matter density with the Belle~II excess.

\subsection{Cosmology during reheating}
\label{sec:reheating}

Following the end of inflation, we consider an inflaton field $\phi$ oscillating at the bottom of a quadratic potential $V(\phi) \propto \phi^2$. During these coherent oscillations, the time averaged equation of state parameter vanishes ($ w_\phi = 0$), causing the inflaton energy density to behave as pressureless matter~\cite{Turner:1983he}. The energy density of the Universe is initially dominated by the inflaton, which progressively transfers its energy into SM radiation through its perturbative decay.\footnote{The field $\phi$ can also decay directly into ALPs or DM; however, this contribution can be neglected as long as the corresponding branching ratios are small enough. For inflatons decaying into DM, it suffices to have ${\rm Br}(\phi \to \chi \chi)\lesssim10^{-4}\,m_\chi/(100\GeV)$~\cite{Drees:2017iod,Arias:2019uol}.}

The evolution of the inflaton energy density $\rho_\phi$ and the SM radiation energy density $\rho_R$ is governed by the coupled Boltzmann equations:
\begin{align}
    \frac{d\rho_\phi}{dt} + 3\, H\, \rho_\phi &= - \Gamma_\phi\, \rho_\phi \,, \label{eq:boltz_phi} \\
    \frac{d\rho_R}{dt} + 3(1+\omega_R) \, H \, \rho_R &= + \Gamma_\phi\, \rho_\phi \,, \label{eq:boltz_R}
\end{align}
where $\Gamma_\phi$ denotes the total decay width of the inflaton into SM radiation, taken to be constant, $H \equiv \dot{a}/a$ is the Hubble parameter, and $\omega_R$ is equation of state parameter for radiation, usually taken equal to $1/3$. The total expansion rate of the Universe is determined by the Friedmann equation,
\be
    H^2 = \frac{\rho_\phi + \rho_R}{3\, M_P^2} \,,
\label{eq:hub}
\ee
with $M_P \simeq 2.4 \times 10^{18}\GeV$ being the reduced Planck mass. The temperature $T$ of the SM thermal bath is related to the corresponding radiation energy density through
\be
    \rho_R(T) = \frac{\pi^2}{30}\, g_*(T)\, T^4 \,,
\label{eq:rad_temp}
\ee
with $g_*(T)$ representing the effective number of relativistic degrees of freedom contributing to $\rho_R$~\cite{Drees:2015exa}.

Throughout this era, the continuous decay of the inflaton injects entropy into the SM thermal bath until $a = a_{\text{rh}}$, defined by $\rho_\phi(a_{\text{rh}}) = \rho_R(a_{\text{rh}})$. The so-called reheating temperature is then defined as $T_{\text{rh}} \equiv T(a_{\text{rh}})$, marking the onset of the standard radiation-dominated era. In order to preserve the successful predictions of BBN, reheating must be completed before the latter starts, which requires $T_{\text{rh}} \gtrsim 4\,\MeV$~\cite{Sarkar:1995dd,Kawasaki:2000en,Hannestad:2004px,DeBernardis:2008zz,deSalas:2015glj}.

To solve Eqs.~\eqref{eq:boltz_phi} and~\eqref{eq:boltz_R}, we follow the procedure described in Ref.~\cite{Silva-Malpartida:2023yks}.\footnote{This approach considers the initial condition $\rho_R=0$, corresponding to a Universe initially dominated by the inflaton. A different cosmological evolution arises if a pre-existing radiation component is present, with $\rho_R>\rho_\phi$ initially; this scenario and its impact on the subsequent reheating dynamics are discussed in detail in Ref.~\cite{Silva-Malpartida:2024emu,Bernal:2025szh}.} An important refinement with respect to the latter work is that here we take into account the temperature dependence of the equation of state parameter, $\omega_R=\omega_R(T)$, while the latter analysis considered a constant $\omega_R=1/3$. This modification becomes particularly relevant for light DM particles, which freeze out at temperatures around the QCD transition ($T \approx 100\,\MeV$).

The entropy injection from the decay of the $\phi$ field modifies the DM freeze-out dynamics and, consequently, the relic abundance and viable parameter space. These effects are discussed in the next Section.

\subsection{Impact on DM production}
\label{sec:dmgenesis}

In this work we consider a thermal Majorana DM candidate whose relic abundance is determined through the freeze-out mechanism. Since $\chi$ is self-conjugate, its cosmological evolution is described in terms of a single number density $n_\chi$, governed by the Boltzmann equation~\cite{Holst:2021lzm}
\be
    \frac{dn_\chi}{dt}+3Hn_\chi
    =
    -\langle\sigma v\rangle
    \left(n_\chi^2-n_{\mathrm{eq}}^2\right),
\label{eq:boltzdm}
\ee
where $n_{\mathrm{eq}}$ denotes the equilibrium DM number density and $\langle\sigma v\rangle$ is the thermally averaged annihilation cross section. Under Maxwell--Boltzmann statistics, the corresponding
$\langle\sigma v\rangle$ 
for a generic $2\to2$ process is given by 
\begin{equation} \label{sigv}
    \langle\sigma\, v\rangle = \int_{4m_\chi^2}^{\infty} \mathrm{d}s\, \frac{(s-4m_\chi^2) \, \sqrt{s} \, K_{1}\left ( \sqrt{s}/T \right ) \sigma(s)}{8 \,T\, m_\chi^{4}\, K_2^2\left ( m_\chi/T \right ) }\,,
\end{equation}
where $K_i$ is the modified Bessel function. For DM production, according to the model described in Sec.~\ref{sec:model}, we consider the $s$-channel annihilation process $\chi\chi \to a^*\to\gamma\gamma$ mediated by an ALP. The $\chi\chi \to aa$ channel is also allowed in the model, along with other ALP-mediated $s$-channel annihilations with SM final states. However, regarding the former, the benchmark scenarios most interesting for our analysis take $m_a>m_\chi$, rendering this process kinematically inaccessible at the freeze-out temperature $T_{\rm fo}$. With respect to the latter, annihilations into heavy SM bosons are also kinematically inaccessible at $T_{\rm fo}$, while annihilations into fermions are loop suppressed, such that they can be safely ignored.

The diphoton annihilation process can be calculated with~\cite{Ding:2025eqq,Dolan:2017osp}: 
\be
    \sigma (s) =\frac{2g^2_{a\chi\chi}\,g^2_{a\gamma\gamma}\, m_\chi^2\, s^{5/2}}{\pi\sqrt{s-4m_\chi^2} \, \left [ m_a^2\,\Gamma_a^2+(m_a^2-s)^2 \right ]}
\label{eq:sigmaV}
\ee
where $s$ is the centre-of-mass energy ($\approx4m_\chi^2$ if non-relativistic) and $\Gamma_a$ is the total width of the ALP. We have checked the validity of this formula numerically using the \texttt{standalone} subroutine from \texttt{MadGraph}~\cite{Alwall:2014hca}. The corresponding \texttt{UFO} model was generated using \texttt{FeynRules}~\cite{Alloul:2013bka}, by adapting the implementation in~\cite{Brivio:2017ije} to the Lagrangian defined by Eqs.~\eqref{eq:LEW} and~\eqref{eq:LDM}.


\begin{figure}[t]
    \centering
    \includegraphics[width=0.49\textwidth]{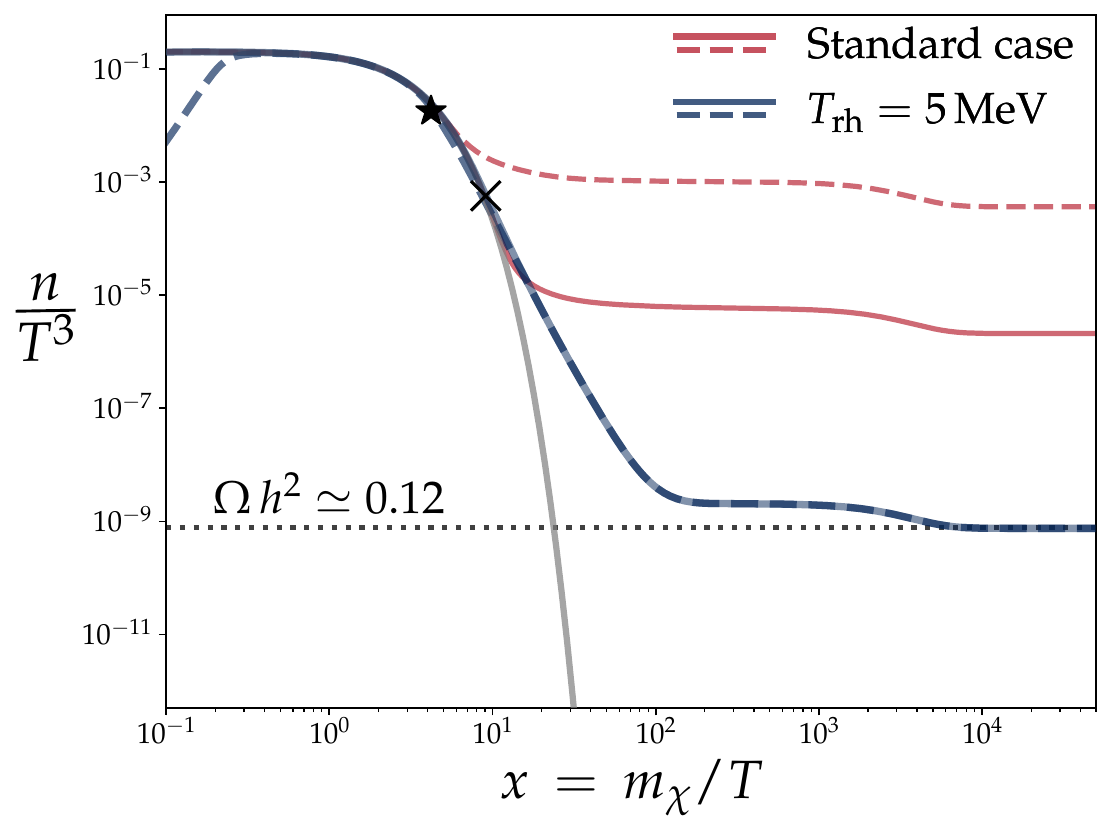}
    \includegraphics[width=0.49\textwidth]{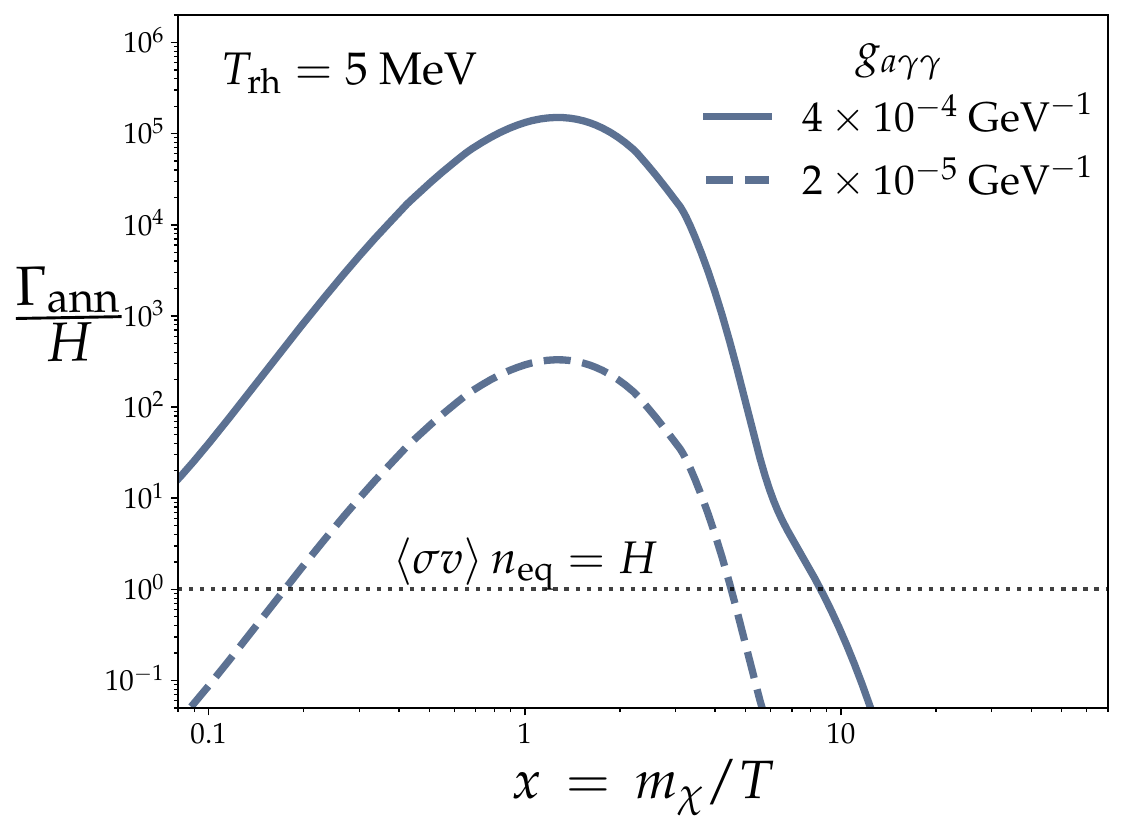}
   \caption{\textit{Left:} Evolution of the normalised DM number density $n_\chi/T^3$ as a function of $x=m_\chi/T$. Red curves correspond to standard cosmology, while blue curves show the scenario with $T_{\rm rh}=5~{\rm MeV}$. In both cases, the solid (dashed) curves correspond to $g_{a\gamma\gamma}=4\times10^{-4}~{\rm GeV}^{-1}$ ($2\times10^{-5}~{\rm GeV}^{-1}$). The grey solid line denotes the equilibrium number density, while the horizontal dotted line indicates $\Omega_\chi h^2\simeq0.12$. \textit{Right:} Ratio $\Gamma_{\rm ann}/H=\langle\sigma v\rangle n_{\rm eq}/H$ for the same two couplings, considering only the $T_{\rm rh}=5~{\rm MeV}$ scenario, with the same solid and dashed line conventions. The horizontal dotted line marks $\langle\sigma v\rangle n_{\rm eq}=H$. Both panels use $m_\chi=500~{\rm MeV}$, $m_a=2.5~{\rm GeV}$, and $g_{a\chi\chi}=0.1~{\rm GeV}^{-1}$.}
    \label{fig:numberDM}
\end{figure}
The effects of entropy injection into the radiation bath on the DM evolution are shown on the left panel of Fig.~\ref{fig:numberDM}. We consider the benchmark point $m_\chi=500\MeV$, $m_a=2.5\GeV$, and $g_{a\chi\chi}=0.1\GeV^{-1}$, and show the evolution of the normalised DM number density $n_\chi/T^3$ as a function of $x\equiv m_\chi/T$. The red and blue curves correspond to standard cosmology and a scenario with $T_{\rm rh}=5\MeV$, respectively, while solid (dashed) lines denote $g_{a\gamma\gamma}=4\times10^{-4}\GeV^{-1}$ ($2\times10^{-5}\GeV^{-1}$).

In the low reheating scenario, the decay of the $\phi$ field into SM radiation continuously injects entropy into the thermal bath, keeping it hotter than in the standard adiabatic evolution at a given scale factor. As a consequence, the temperature decreases more slowly and $n_\chi/T^3$ is suppressed, as shown by the blue curves. Therefore, these scenarios require a $\langle\sigma\,v\rangle$ that in standard cosmology would produce an overabundance (compare blue and red curves), in other words, when one has freeze-out with a low reheating temperature, one generally reproduces the observed relic abundace $\Omega h^2$ with smaller couplings than in standard cosmology.\footnote{In addition to the entropy injection, the Hubble constant is also modified due to the dominance of $\rho_\phi$ over $\rho_R$ (see Eq.~\eqref{eq:hub}). Contrary to the aforementioned dilution, this effect favours larger couplings. However, as shown in Ref.~\cite{Bernal:2018kcw}, the entropy injection has a larger impact on the resulting relic abundance, meaning that an overall smaller coupling is preferred.}

An interesting feature of our model is that it involves light DM. In combination with the weaker couplings required for $\Omega h^2$, this drives an earlier decoupling from equilibrium. As shown on the right panel of Fig.~\ref{fig:numberDM}, this is determined by $\langle\sigma\,v\rangle\,n_{\rm eq}/H=1$, such that the considered benchmark points have $x_{\rm fo}\simeq10$ and~4, for $g_{a\gamma\gamma}=4\times10^{-4}\GeV^{-1}$ and $2\times10^{-5}\GeV^{-1}$, respectively. The corresponding points are indicated on the left panel by $``\times"$ and $``\star"$ symbols. Indeed, in most of our results we will find small values of $x_{\rm fo}$, with some cases even displaying decoupling while DM is still relativistic. Having warm, or hot, DM usually is in conflict with large scale structure formation, however, in cosmological scenarios with late reheating this is not necessarily the case.  As shown in Refs.~\cite{Henrich:2025sli,Henrich:2025gsd}, if freeze-out happens during the reheating epoch, the DM population undergoes sufficient subsequent cooling, such that it becomes cold by the time of structure formation, meaning these scenarios are still acceptable.\footnote{The relevant constraints arise from the free-streaming suppression of small-scale structure, which can be probed through Lyman-\(\alpha\) observations \cite{Viel:2013fqw,Irsic:2017ixq}, leading to the constraint $m_\chi>5~{\rm keV}\,(T_{\rm rh}/T_{\rm fo})^{5/3}$.}

Another interesting feature of our setup is that, for a fixed $T_{\rm rh}$, smaller couplings not always lead to larger abundances. As can be seen on the left panel of Fig.~\ref{fig:numberDM}, two very different couplings lead to exactly the same $\Omega h^2$. The reason for this is that the $n_{\rm eq}$ curve has a different height and slope at the moment where each decoupling occurs, such that the posterior evolution of $n/T^3$ in each case ends up coinciding. Thus, in our final results shown in the next Section, we shall find that for each $T_{\rm rh}$ we will find two sets of solutions giving the correct relic abundance, each corresponding to very different $x_{\rm fo}$.

\subsection{Compatibility of DM genesis and Belle~II excess}
\label{sec:cosmo-compat}

\begin{figure}[tb]
\centering
\includegraphics[width=\textwidth]{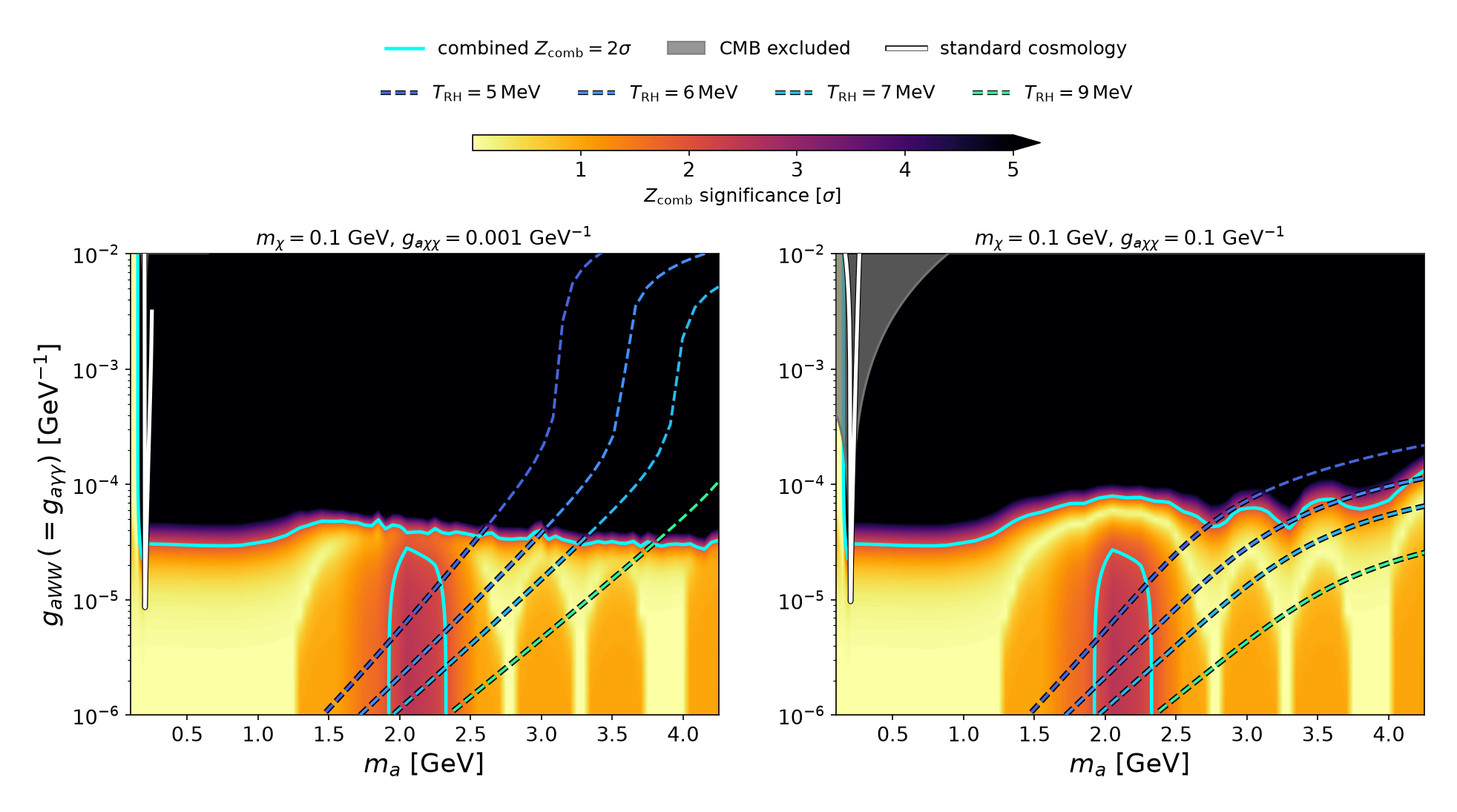}
\caption{\it Combined exclusion significance $Z_{\rm comb}$ of Eq.~\eqref{eq:combined-exclusion}
(colour map and cyan $2\sigma$ contour) in the $(m_a,g_{aWW})$ plane
, for $m_\chi=0.1\GeV$ and $g_{a\chi\chi}=10^{-3},\,10^{-1}\GeV^{-1}$ (left to right). The solid
white line is the standard-cosmology relic-density solution; dashed lines show the same solution
once entropy injection from a low reheating temperature is included (for $T_{\rm rh}=5,\,6,\,7,\,9$~MeV).
The grey region is excluded by the cosmic microwave background (CMB) bound on late dark-matter energy
injection.}
\label{fig:relic-overlay-mchi01}
\end{figure}

\begin{figure}[tb]
\centering
\includegraphics[width=0.89\textwidth]{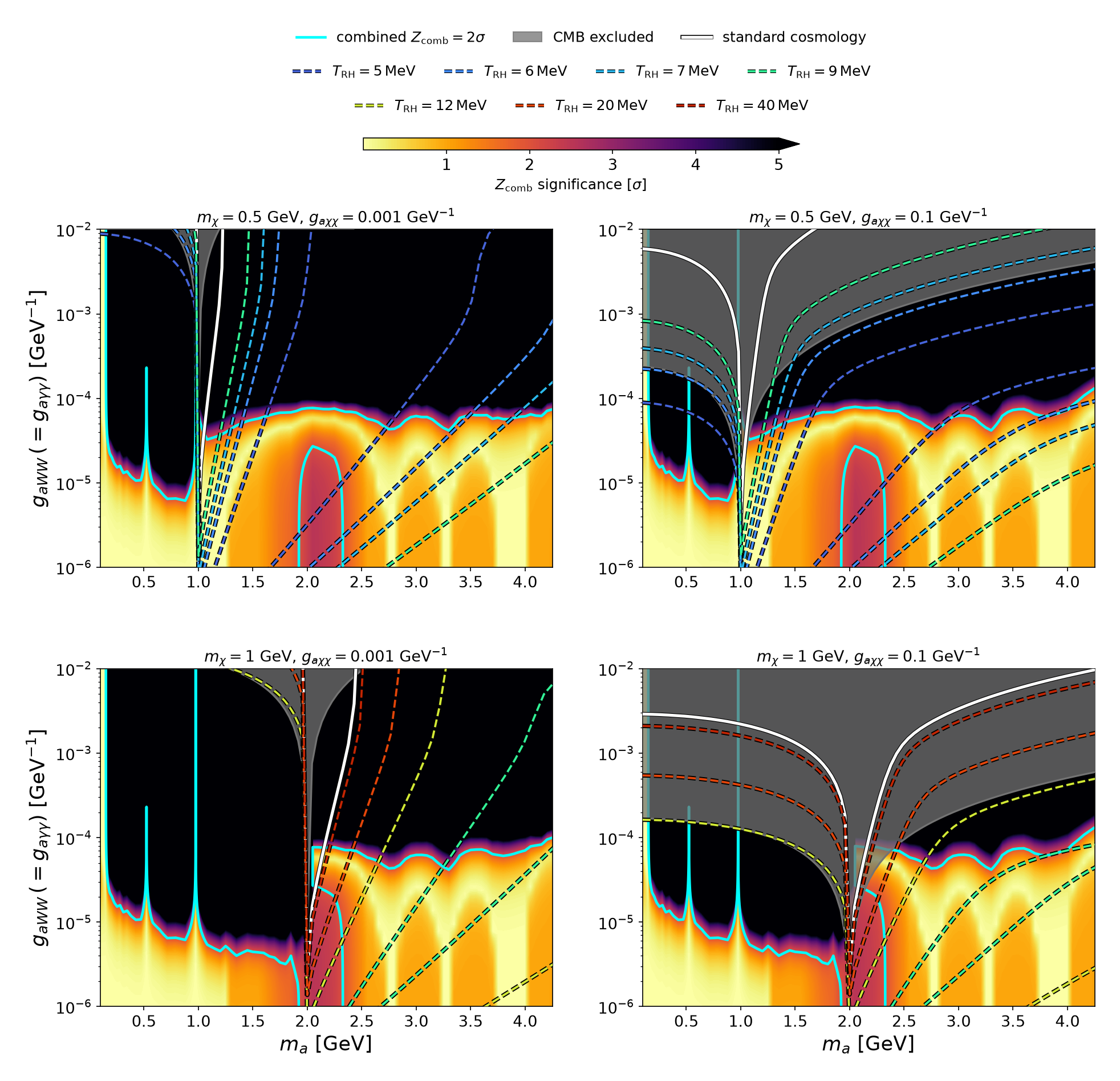}
\caption{\it Same as Fig.~\ref{fig:relic-overlay-mchi01}, for $m_\chi=0.5\GeV$ (top row) and
$m_\chi=1\GeV$ (bottom row). To keep the panels legible, only a subset of the available
$T_{\rm rh}$ values is shown: $5\,,6\,,7\,,\,9$~MeV for $m_\chi=0.5\GeV$, and $9,\,12,\,20,\,40$~MeV for
$m_\chi=1\GeV$.}
\label{fig:relic-overlay-mchi05-1}
\end{figure}

We now examine whether a low reheating temperature reconciles our ALP-mediated dark matter model
with the Belle~II excess, following our benchmark of Eq.~(\ref{eq:benchmark}). Figs.~\ref{fig:relic-overlay-mchi01} and~\ref{fig:relic-overlay-mchi05-1}
show the same information from Figs.~\ref{fig:combined-terrestrial-mchi01} and~\ref{fig:combined-terrestrial-mchi05-1}, 
this time combining all bounds from the flavour sector via $Z_{\rm comb}$,
overlaid with curves reproducing the observed dark matter abundance. The
solid white lines show the solution for standard cosmology.\footnote{Note that in Ref.~\cite{Ding:2025eqq} the effects of an early kinetic decoupling are included. We neglect these effects, as Ref.~\cite{Ding:2025eqq} found them to matter only in the vicinity of the resonance ($2m_\chi \lesssim m_a \lesssim 2.7\,m_\chi$). Our solutions are usually far from the funnel, so we expect them to be generally unaffected. For curves close to $m_a \simeq 2m_\chi$ the required coupling may be shifted upwards by a factor of a few.} There, the observed relic abundance is
reproduced within the resonant funnel around $m_a\approx2m_\chi$, where the $s$-channel process
$\chi\chi\to a^*\to\gamma\gamma$ is resonantly enhanced. This solution is not excluded. It is, however,
compatible with flavour data only over a narrow range of $m_a$, usually well below the
$m_a\approx2\GeV$ values relevant for explaining the Belle~II excess.

In addition to this, on all plots we also overlay as dashed lines the corresponding solutions once entropy injection from a low reheating temperature is taken into account. These solutions allow for compatibility with flavour data over broader ranges and at larger values of $m_a$, which is the main result of this work.

For $m_\chi=0.1\GeV$, Fig.~\ref{fig:relic-overlay-mchi01} shows the relevant region of the
$(m_a,\,g_{aWW})$ plane, with the left and right panels corresponding to $g_{a\chi\chi}=10^{-3}$ and
$0.1\GeV^{-1}$, respectively. The solid white line, the standard cosmology solution, lies at
$m_a\approx0.2\GeV$. The dashed low-reheating lines instead sit at $m_a\gtrsim1.2\GeV$. On the left panel, these extend
to arbitrarily large $g_{aWW}$, while on the right they reach a maximum around
$g_{aWW}\sim5\times10^{-4}\GeV^{-1}$.  For this case, chemical decoupling occurs while the DM is still relativistic, $T_{\rm fo}>m_\chi$, in contrast to the conventional non-relativistic WIMP freeze-out at $x_{\rm fo}\sim\mathcal{O}(20)$. However, as we have argued in the previous Section, this will not be a problem, as the DM will be cold by the time of structure formation.

Let us now describe the behaviour of the dashed lines for the observed relic density, which depend crucially on the decay channel dominating the ALP total width $\Gamma_a$. On the right panel ($g_{a\chi\chi}=0.1~{\rm GeV}^{-1}$), for small $m_a$ and $g_{aWW}$, the width is dominated by the invisible channel, $a\to\chi\chi$, meaning that $\Gamma_a$ will not be sensitive to the variation of $g_{aWW}$, and will depend linearly on $m_a$, to a very good approximation. Thus, in order to keep the annihilation rate in Eq.~\eqref{eq:sigmaV} compatible with the observed relic abundance, an increase in $m_a$ must be compensated by a larger $g_{aWW}$. This sets the slope of the relic density contours in this region. For larger values of $g_{aWW}$ and $m_a$, the decay $a\to\gamma\gamma$ dominates the total ALP width instead. Since $\Gamma_a$ now depends on $g_{aWW}$ and $m_a^3$, this leads to a modification in the slope of the dashed lines. 
These curves shift to the right when we increase $T_{\rm rh}$ since here the decoupling always happens before the Boltzmann suppression has set it. Thus, since a larger $T_{\rm rh}$ implies a smaller entropy injection, the decoupling needs to happen slightly later in the nearly constant $n_{\rm eq}/T^3$ curve. In other words, for this set of curves, increasing $T_{\rm rh}$ requires a larger $\langle\sigma v\rangle$.

On the left panel ($g_{a\chi\chi}=10^{-3}~{\rm GeV}^{-1}$) we find the same situation as on the right panel when $m_a$ and $g_{aWW}$ are small, and when $m_a$ and $g_{aWW}$ are large. However, we find that both regimes are connected by intermediate nearly vertical dashed lines, which deserve an explanation, regardless of the fact that they do not solve the Belle~II excess. Here, we find that the small value of $g_{a\chi\chi}$ leads to an even larger $T_{\rm fo}$ than on the right panel. This means that the integrand found in $\langle\sigma\, v\rangle$ (see Eq.~\eqref{sigv}) is not necessarily Boltzmann suppressed at relatively large values of $s$. This allows the integral to be dominated by an ALP funnel at $s=m_a^2$, regardless of the fact that $m_a> 2m_\chi$, so one can demonstrate that $\langle\sigma \,v\rangle\propto g_{a\chi\chi}^2 \,g_{a\gamma\gamma}^2/(m_a\Gamma_a)$ (see also the discussion in Ref.~\cite{Gondolo:1990dk}). This result does not depend on $g_{aWW}$ in the limit of $a\to\gamma\gamma$ dominating $\Gamma_a$, explaining thus the vertical lines. At very large $g_{aWW}$ the value of $T_{\rm fo}$ is lower, so the integration over $s$ is Boltzmann suppressed at the ALP funnel, leading to the same slope as in the right panel.

Fig.~\ref{fig:relic-overlay-mchi05-1} extends this analysis to two heavier benchmarks, $m_\chi = 0.5\GeV$ (top row) and $m_\chi = 1\GeV$ (bottom row), again for $g_{a\chi\chi} = 10^{-3}\GeV^{-1}$ (left) and $0.1\GeV^{-1}$ (right). In both cases, chemical decoupling instead occurs while the dark matter is already non-relativistic, but still feature small values of $x_{\rm fo}$, specially at large $m_a$ and small $g_{aWW}$.

As the DM mass increases, important changes happen in the relic density contours. First, the resonant funnel zone for standard cosmology (solid white line) shifts towards larger ALP masses following the condition $m_a \simeq 2m_\chi$. Second, for fixed couplings, the increase in $m_\chi$ makes the annihilation process $\chi\chi\to a^*\to\gamma\gamma$ become more efficient, as a consequence of the mass dependence induced by the coupling between $\chi$ and the ALP. This lowers the abundance prior to dilution. Since the effect is compensated by lowering $g_{aWW}$, the dashed lines on the right side of the plot, analogous to those in Fig.~\ref{fig:relic-overlay-mchi01}, are shifted towards smaller values of the latter. Notice, however, that this is not convenient, as the lines no longer pass through the $m_a\approx2$~GeV region for the Belle~II excess.

In addition to the aforementioned effects, new dashed lines can be seen around the resonant funnel. These correspond to the second set of solutions mentioned in Sec.~\ref{sec:dmgenesis} for larger $x_{\rm fo}$, where Boltzmann suppression is stronger. Again, since entropy injection during reheating reduces the final DM abundance, matching the observed relic density requires a smaller $\langle\sigma \,v\rangle$ compared to standard cosmology. This is achieved through a reduction of $g_{aWW}$ and a shift of $m_a$ away from the resonant condition $m_a\simeq2m_\chi$. However, in contrast to the first set of solutions, for this set an increase in $T_{\rm rh}$ implies smaller $\langle\sigma\, v\rangle$, and so a larger separation from the resonant condition. Furthermore, as $\langle\sigma\, v\rangle$ depends on the DM mass, we find this effect is also affected by $m_\chi$. For $m_\chi=0.5\GeV$ it produces only a moderate modification near the funnel. For $m_\chi=0.1\GeV$ it is not visible at all, as the curves for different reheating temperatures closely overlap with the standard cosmology prediction. For $m_\chi=1\GeV$, the larger annihilation efficiency leads to a significant departure from the resonant condition for values of $T_{\rm rh}$ as large at 10~MeV. This broadening of the funnel is therefore most visible for large DM masses.

On the upper panels of Fig.~\ref{fig:relic-overlay-mchi05-1}, for $m_{\chi}=0.5$~GeV, we find a gap between the first and second set of lines. For this example, it is unfortunate that this gap coincides with $m_a\approx2$~GeV region associated to the Belle~II excess. This solution would predict that future flavour data should shift the preferred region to larger values of $m_a$. However, as will be seen below, setting $c_b\neq c_W$ can allow one set of lines to coincide with the favoured region.

In contrast, the lower panels of Fig.~\ref{fig:relic-overlay-mchi05-1}, for $m_\chi=1$~GeV, show no such gap between the sets of lines, with a minimum $T_{\rm rh}\approx9$~MeV.
Notice that here the solution for standard cosmology is compatible with the $m_a\approx2$~GeV excess, with a clear conflict with the CMB constraint when setting $g_{a\chi\chi}=0.1$. We find that the role of the late reheating in this case is to help avoid the CMB bounds, with a solution allowed by the latter and passing through the $m_a\approx2$~GeV region for $T_{\rm rh}\approx 20$~MeV.

\begin{figure}[tb]
\centering
\includegraphics[width=\textwidth]{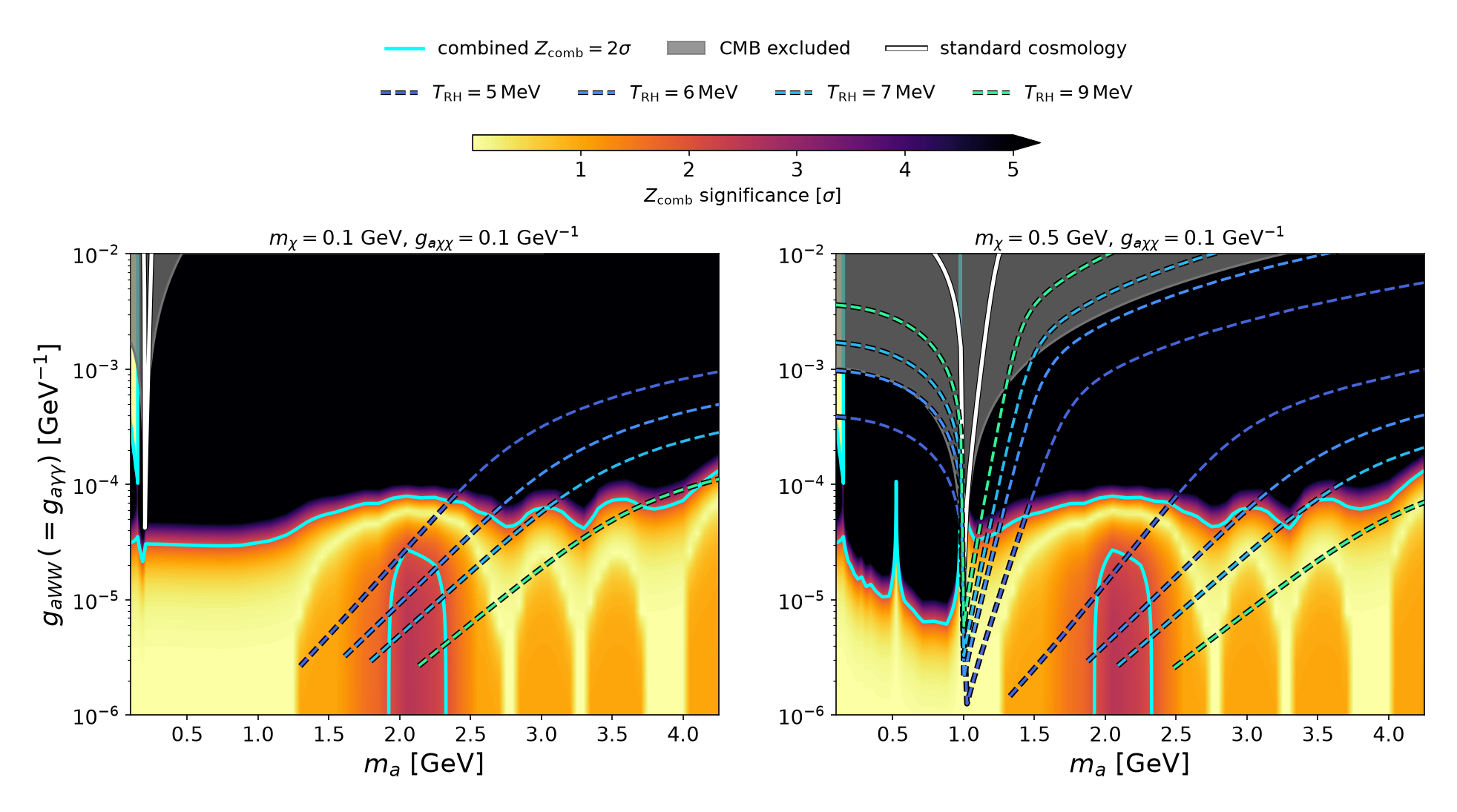}
\caption{\it Same as Figs.~\ref{fig:relic-overlay-mchi01} and~\ref{fig:relic-overlay-mchi05-1}, for
$\rho_{\rm EW}=0.23$ ($g_{a\gamma\gamma}=0.23\,g_{aWW}$) and a single $g_{a\chi\chi}=0.1\GeV^{-1}$,
for $m_\chi=0.1$ (left) and $0.5\GeV$ (right). As in
Figs.~\ref{fig:relic-overlay-mchi01} and~\ref{fig:relic-overlay-mchi05-1}, only a sample
of the available $T_{\rm rh}$ values is shown: $5,\,6,\,7,\,9\,$MeV. For very small $g_{aWW}$ the dashed lines are cut off, as DM does not reach thermal equilibrium.}
\label{fig:relic-overlay-rhoEW}
\end{figure}
Let us now consider the situation where $g_{a\gamma\gamma}\neq g_{aWW}$, addressed in Section~\ref{sec:rhoEW-outlook} via the parameter $\rho_{\rm EW}$. The relic abundance overlay of Figs.~\ref{fig:relic-overlay-mchi01}
and~\ref{fig:relic-overlay-mchi05-1} responds to $\rho_{\rm EW}$ more simply than the flavour
bounds of Sec.~\ref{sec:rhoEW-outlook}. Both the annihilation rate $\chi\chi\to a^*\to\gamma\gamma$
of Eq.~\eqref{eq:sigmaV} and the late dark-matter energy injection constrained by the CMB bound
are controlled by $g_{a\gamma\gamma}=g_{aWW}\,\rho_{\rm EW}$ alone. Therefore, the standard cosmology line,
every low-reheating branch, and the CMB-excluded grey band all rescale rigidly as
$g_{aWW}\to g_{aWW}/\rho_{\rm EW}$, uniformly across all six panels: up by a factor $\simeq4.3$ at
$\rho_{\rm EW}=0.23$ and down by a factor $5$ at $\rho_{\rm EW}=5$. The main result of this work, namely that a low reheating temperature reproduces the observed relic density over a wider range of $m_a$ than standard cosmology, therefore survives unchanged at both values of $\rho_{\rm EW}$. Nevertheless, as can be seen in Fig.~\ref{fig:relic-overlay-rhoEW}, the shift induced by the small values of $\rho_{EW}=0.23$ allows an even greater compatibility between the low $T_{\rm rh}$ solutions and the Belle~II excess at $m_a\approx2$~GeV. We show this only for $m_\chi=0.1$ and $0.5\GeV$.\footnote{At $m_\chi=1\GeV$, standard cosmology already reproduces the excess without requiring $\rho_{\rm EW}\neq1$, as noted above. This benchmark therefore carries no new information under this variation.} The left panel of the Figure shows the situation for $m_\chi=0.1$~GeV and $g_{a\chi\chi}=0.1~{\rm GeV}^{-1}$ (to be compared with the right panel of Fig.~\ref{fig:relic-overlay-mchi01}), where we see that a larger number of dashed lines cross the $m_a\approx2$~GeV region. Moreover, on the right panel we have $m_\chi=0.5$~GeV and $g_{a\chi\chi}=0.1~{\rm GeV}^{-1}$ (to be compared with the upper right panel of Fig.~\ref{fig:relic-overlay-mchi05-1}). Here we find that the gap between the two sets of solutions is now displaced to a different range of $m_a$, such that now one can find dashed lines from one set of solutions crossing the $m_a\approx2$~GeV region. Similarly, one finds that setting $\rho_{\rm EW}=5$ allows lines from the other set of solutions to cross this region.


\section{Conclusions}
\label{sec:Conclusions}

This work set out to answer one question: if an ALP mediates both the Belle~II excess in
$B^+\to K^++{\rm inv}$ and the annihilation of a thermal dark-matter candidate $\chi$, does
reproducing the observed relic density require abandoning the standard cosmological history? 

On the first part of this study, in Sec.~\ref{sec:alpportal}, we considered an update on the analysis of the excess, based on a mass-differential Belle~II
likelihood~\cite{Abumusabh:2025zsr} combined with LHCb and Belle visible-channel searches that earlier analyses omitted. This was confronted with Ref.~\cite{Ding:2025eqq}, which considered this scenario in the context of standard cosmology, and found that their resonant-annihilation benchmarks were excluded, at $Z_{\rm comb}>2.8$.
In particular, overlaying our own relic density solutions on the
same exclusion map in Sec.~\ref{sec:cosmo}, we find that under a standard thermal history the
resonant condition $m_a\simeq2m_\chi$ survives flavour and collider data only in a narrow window,
well below the ALP masses the excess favours.

A low reheating temperature removes this tension. Entropy injected as the inflaton decays dilutes
the abundance that freeze-out would otherwise produce, so a smaller annihilation rate, and hence a
smaller ALP coupling, is enough to match $\Omega_{\rm obs}h^2$. This broadens the viable region and
shifts it to larger $m_a$, up to $\gtrsim1.2\GeV$ at the lightest benchmark studied and further
still as $m_\chi$ grows. The shift moves the viable region into the $m_a\approx2$~GeV mass range the excess points
to, and into couplings the current flavour bounds leave open. A reheating temperature as low as
the $T_{\rm rh}\gtrsim4\,\MeV$ floor allowed by BBN is enough to achieve this. 

\textit{The take-home message is that, in this scenario, the Belle~II excess and a thermal dark-matter
origin are not simultaneously viable under a standard cosmology, at least not at the masses the
excess favours. What reconciles them is a change in the pre-BBN history rather than in the ALP
couplings.} Because the coupling window this opens is within the projected reach of a future linear-collider measurement~\cite{Bhattacharya:2025hme}, the scenario remains falsifiable. Confirming or excluding it will also constrain what the Universe was doing before BBN.

\section*{Acknowledgements}

We are grateful to Marta Fuentes Zamoro for valuable discussions. L.~M.~acknowledges the Secci\'on F\'isica of the Departamento de Ciencias at the Pontificia Universidad Cat\'olica del Per\'u while J.~S.~M. acknowledges the Instituto de Física Teórica of the Universidad Autónoma de Madrid for hospitality while this project was carried out.

J.~J.~P.~and J.~S.~M.~acknowledge funding by the \textit{Dirección de Gestión de la Investigación} at PUCP, through grant DFI-PUCP-PI1144. L.~M.~acknowledges partial financial support by the Spanish Research Agency (Agencia Estatal de
Investigaci\'on) through the grant IFT Centro de Excelencia Severo Ochoa No CEX2025-001574-S (within the research line Particle Physics in the Standard Model and Beyond, BSM) and by the grants PID2022-137127NB-I00, PID2025-172338NB-I00 and PID2025-171019NB-C21 funded by MICIU/AEI/10.13039/501100011033.

This project has received funding from the European Union’s Horizon Europe research and innovation programme under the Marie Skłodowska-Curie Staff Exchange grant agreement No 101086085 – ASYMMETRY.


\bibliographystyle{utphys}
\bibliography{references}

\end{document}